\documentclass[final,nomarks]{dmtcs-episciences}

\usepackage[utf8]{inputenc}
\usepackage{amsmath}
\usepackage{stmaryrd}
\usepackage{amssymb}

\usepackage{subcaption} 

\usepackage{tikz}
\usepackage{tikz-cd}
\usetikzlibrary{automata,positioning,arrows}

\newcommand{\Z}{\integers}
\newcommand{\Zmod}[1]{\Z_{#1}}

\newcommand{\N}{\naturals}
\newcommand{\joliN}{\mathcal{N}}
\newcommand{\supp}{\operatorname{supp}}

\newcommand{\id}{\operatorname{id}}
\newcommand{\lcm}{\operatorname{lcm}}
\newcommand{\intint}[1]{\left\llbracket #1 \right\rrbracket}
\newcommand{\iindex}{\mathfrak{t}}
\newcommand{\period}{\mathfrak{p}}
\newcommand{\zero}[1]{\star_{#1}}
\newcommand{\multi}[1]{\mathbf{#1}}
\newcommand{\floor}[1]{\left\lfloor#1\right\rfloor}
\newcommand{\freemonoid}[2]{C_{#1,#2}}

\newtheorem{prop}{Proposition}
\newtheorem{defi}{Definition}
\newtheorem{thm}{Theorem}
\newtheorem{lemma}{Lemma}

\usepackage[round]{natbib}

\author[Vincent Nesme]{Vincent Nesme\affiliationmark{1}}
\title[Automaticity and CA on commutative monoids]{Automaticity of spacetime diagrams generated by cellular automata on commutative monoids}
\affiliation{PhD graduate from the \'Ecole normale supérieure de Lyon, France; lycée Eugène Delacroix, Maisons-Alfort, France}
\begin{document}
\publicationdata{vol. 28:3}{2026}{8}{10.46298/dmtcs.9846}{2022-07-27; 2022-07-27; 2025-07-08; 2026-02-17}{2026-08-07}

\maketitle


\begin{abstract}
\vspace{0.2cm}
It is well known that the spacetime diagrams of some cellular automata have a fractal structure: for instance, Pascal's triangle modulo 2 generates a Sierpi\'nski triangle.
It has been shown that such patterns can occur when the alphabet is endowed with the structure of an abelian group, provided the cellular automaton is a morphism with respect to this structure and the initial configuration has finite support.  The spacetime diagram then has a property related to $k$-automaticity.
We show that these conditions can be relaxed: the abelian group can be a commutative monoid, the initial configuration can be $k$-automatic, and the spacetime diagrams still exhibit the same regularity. 
\end{abstract}

\section*{Introduction}

This work is part of a series of articles aiming at classifying cellular automata into meaningful subsets.  Our starting point here is the well-known fact that Pascal's triangle modulo 2, which can be computed by a simple cellular automaton performing an XOR, produces a spacetime diagram that converges to a Sierpi\'nski triangle.
From there, a series of questions emerges.  Why? How does it work? Can we characterize a class of cellular automata that exhibit similar behaviors? 

Some authors have studied the graphical limit sets of cellular automata with very lax algebraic structures, or no structure at all --- see for instance \cite{haeseler1993cellular,haeseler2001self1,haeseler2001self2,muller2015graphical}.   We shall impose a strong algebraic constraint on the transition rule of the cellular automaton, 
and look at what can be deduced about its spacetime diagram.  Our long-term objective is to discuss ``summable cellular automata", for which it makes sense to isolate the influence of a single cell, and where the global transition function can be reconstructed by ``summing" all these influences.  Predicting the state of a cell in such a cellular automaton is expected to be an easy task, since no interaction is allowed to take place, but actually finding a description of the spacetime diagram is a nontrivial task.

Let $\Sigma$ denote the alphabet.  Instead of the usual local transition function $\Sigma^{\mathcal{N}}\to \Sigma$, a summable cellular automaton is naturally defined by a function $\Sigma\to\Sigma^{I}$ that describes the influence of each cell on its neighborhood.  Arguably, the minimal algebraic structure allowing us to do that is to endow $\Sigma$ with a binary operation $\cdot$ that makes $(\Sigma, \cdot)$ a commutative monoid, and to require that the cellular automaton of interest be an endomorphism of $(\Sigma,\cdot)^{\Z}$.

One could ask whether we really do need an identity element in $\Sigma$. What if $(\Sigma,\cdot)$ is just a commutative semigroup?  In that case, the algebraic structure is too lax.  Consider an arbitrary cellular automaton on an alphabet $\Sigma$ bearing no algebraic structure.  We can add an element $\star$ to $\Sigma$, and define a binary operation on $\Sigma^\star= \Sigma\cup \{\star\}$ by: for every $x,y\in\Sigma^\star$, $x\cdot y = \star$.  $(\Sigma^\star,\cdot)$ is a semigroup. Now, if $F:\Sigma^\Z\to \Sigma^\Z$ is a cellular automaton, extend it to a cellular automaton $F^\star:(\Sigma^\star)^\Z \to (\Sigma^\star)^\Z $ so that $\star$ is a quiescent state (i.e., the configuration $(\ldots,\star,\star,\star,\ldots)$ is mapped to itself).  Then $F^\star$ is an endomorphism of $(\Sigma^\star,\cdot)^\Z$, but it behaves like the arbitrary cellular automaton $F$ on initial configurations not containing the letter $\star$.  This shows that the dynamics of any arbitrary cellular automaton can be replicated by a cellular automaton that is also a commutative semigroup endomorphism.  It follows that, to obtain nontrivial results about the dynamics of a cellular automaton that is an endomorphism of some algebraic structure, this structure has to be more rigid than a mere commutative semigroup. The next step is adding an identity element to make it a commutative monoid.

Let $(\Sigma,\cdot,1_\Sigma)$ be a finite commutative monoid, $I$ some finite subset of $\Z$, and $(f_i)_{i\in I}$ a family of endomorphisms of $\Sigma$.   We can then define the following endomorphism $F$ of $\Sigma^\Z$:

\begin{equation}\label{eq:globaltransition}
 F: \left(    \begin{array}{rcl}  
\Sigma^\Z & \to & \Sigma^\Z \\
(r_n)_{n\in \Z} & \mapsto  &  \left( \prod\limits_{i\in I}  f_i(r_{n-i}) \right)_{n\in \Z} 
\end{array}
  \right).
\end{equation}

The monoid operation is written multiplicatively throughout most of the paper, except in Section~\ref{sec:groups} and in Lemma~\ref{lemma4}.  
$F$ is a cellular automaton on the alphabet $\Sigma$, with neighborhood included in $-I$.   Conversely, if $F$ is a cellular automaton over the alphabet $\Sigma$ that is also an endomorphism of $\Sigma^\Z$, then one can choose a neighborhood $\joliN$ of $F$, and define, for $i\in I = -\joliN$ and $s\in \Sigma$,
\begin{displaymath}
f_i(s) = F(\bar s)_i ,
\end{displaymath}
where $\bar s$ is the word of $\Sigma^\Z$ defined by $\bar s_n = \left\{  \begin{array}{ll}
s & \text{if $n=0$} \\ 
1_\Sigma & \text{otherwise}
\end{array}\right.,$  $1_\Sigma$ denoting the identity element of $\Sigma$.

The support of a configuration $c\in\Sigma^\Z$ is defined by $\supp(c) = \{n\in \Z ; c_n \neq 1_\Sigma\}$.  We say a configuration is finite if it has finite support.  $\Zmod{m}$ denotes the finite cyclic group of order $m$.

\begin{itemize}
\item The case $(\Sigma,\cdot)=(\Zmod{2}, +)$ was treated by \cite{willson1984cellular}.    It includes Pascal's triangle modulo 2, and describes the fractal structure of the limit spacetime diagram in terms of matrix substitution systems.
\item The case $(\Sigma,\cdot) = (\Zmod{p^k},+)$ was treated by \cite{takahashi1992self}.  It is a generalization of Willson's article.
\item The case $(\Sigma,\cdot)=(\Zmod{m},+)$ was treated by \cite{allouche1997automaticity}.   It describes the spacetime diagram in terms of $k$-automaticity, which is another name for matrix substitution systems, and sorts out for which $k$ the spacetime diagram is $k$-automatic and for which $k$ it is not.  We will also adopt the language of $k$-automaticity in this paper.  
\item The case when $(\Sigma,\cdot)$ is a (finite) abelian group was treated by \cite{Gutschow2010fractal}.  It uses as an example a cellular automaton already studied in \cite{Macfarlane2004linear}.
\end{itemize}

Let us introduce briefly the notions used in the statement of our main results.
A cellular automaton $F$, when running on an initial configuration $c\in\Sigma^\Z$, produces a spacetime diagram $(F^j(c)_i)_{(i,j)\in\Z\times\N}$, that is a double sequence with values in $\Sigma$.  The regularity of such double sequences will be described in terms of $k$-automaticity.   The reference for all things $k$-automaticity is \cite{allouche2003automatic}, in our case particularly its chapter~14, since we are concerned with double sequences.   Actually, we are concerned more specifically with sequences indexed by $(x,y)\in\Z\times\N$ that are, in the language of \cite{allouche2003automatic} and \cite{rowland2020automaticity}, $[-k,k]$-automatic.  Since 
this is the only kind of automaticity we will care about, 
we will write ``$k$-automatic" in lieu of ``$[-k,k]$-automatic".  Here is our definition of $k$-automaticity, which encompasses the usual definitions of $[k,k]$, $[-k,k]$, $[k,-k]$ and $[-k,-k]$-automaticity for double sequences indexed by $\N\times\N$, $\Z\times \N$, $\N\times\Z$ and $\Z\times\Z$.

\begin{defi} \label{def:k-automatic}
Let $d\geq 1$ and $k\geq 2$ be integers.  Let $U$ be a function defined on $D\subseteq \Z^d$. We extend the domain of $U$ by choosing an element  $\bot\not\in U(D)$, and setting $U(x)=\bot$ for any $x\in \Z^d\setminus D$.  $U$ is $k$-automatic if there exists a finite set $E$ and a function $e:\Z^d \to E$ such that 

\begin{itemize}
\item $U(\multi{n})$ is a function of $e(\multi{n})$;
\item for $\multi{s}\in \intint{0,k-1}^d$ and $\multi{n}\in \Z^d$, $e(k\multi{n}+\multi{s})$ is a function of $\multi{s}$  and $e(\multi{n})$.
\end{itemize}
\end{defi}

Notice that, in this definition, $d$ need not be specified: a function defined on $D\subseteq \Z^d$ is $k$-automatic if and only if it is $k$-automatic as a function defined on a subset of $\Z^{d'}$, for any $d'\geq d$.  It should also be noted that $U(D)$ is necessarily a finite set, since $e(D)$ is a finite set and $U(\multi{n})$ is a function of $e(\multi{n})$.

The Prouhet-Thue-Morse sequence is the prototypical $2$-automatic sequence.  As an example, let us see how it is $2$-automatic according to our definition.
The Prouhet-Thue-Morse sequence can be defined as the only function $U$ defined on $\N$ such that $U(0)=0$ and, for every $n\in\N$:

\begin{displaymath}
\left\{ \begin{array}{l} 
U(2n) = U(n) \\
U(2n+1) = 1-U(n)
\end{array}\right.
\end{displaymath}

So, first, we extend the domain of $U$ by setting $U(n)=\bot$ for every $n<0$.  Let $E=\{\bot,0,1\}$ and $e=U$.  $U(n)$ is then a function of $e(n)$, since $U(n)=e(n)$. 
Moreover, for every $n$ in $\Z$,  $e(2n)=e(n)$ and $e(2n+1)=\left\{ \begin{array}{ll} 
1-e(2n) & \text{if $e(n)\in\{0,1\}$} \\
\bot & \text{if $e(n)=\bot$} \\
\end{array}\right.$.

Therefore, for $s\in\{0,1\}$ and  $n\in\Z$, $e(2n+s)$ is a function of $s$ and $e(n)$. This is how our definition accounts for the fact that the Prouhet-Thue-Morse is $2$-automatic.

In the language of $k$-automaticity, the main result of \cite{Gutschow2010fractal} is:

\begin{thm}\label{thm:GNW10}
If $(\Sigma,\cdot)$ is an abelian $p$-group, then the double sequence generated by a cellular automaton that is also an endomorphism of $\Sigma^\Z$, starting on a finite configuration, is $p$-automatic.
\end{thm}

In order to state our new result, let us introduce a few more notations.
It is well known and ubiquitous that the semilinear subsets of $\N^n$ are exactly those definable in Presburger arithmetic.   The same holds in $\Z^n$, provided the order $<$ is included in Presburger arithmetic.  The following equivalence is stated in \cite{choffrut2010deciding} (Theorems 1.1 and 1.3):

\begin{prop}\label{prop:rational}
Given a subset $X$ of $\Z^n$, the following assertions are equivalent:
\begin{enumerate}
\item $X$ is first-order definable in $\left<\Z ;  + ,<,0,1 \right>$;
\item $X$ is $\N$-semilinear, i.e., it is a finite union of sets of the form $a+\sum\limits_{i=1}^{k} \N b_i$,  where $a,b_i\in \Z^n$;
\item $X$ is a rational subset of $\Z^n$ , i.e., it can be obtained from singletons of $\Z^n$ by applying the union, product and Kleene star operations a finite number of times.
\end{enumerate}
\end{prop}

Of particular use for us will be the corollaries that any boolean combination of rational subsets of $\Z^n$ is rational, and that if $X$ is a rational subset of $\Z^{n}\times \Z$, then $\left\{ (x,y) ; (x,\ldots,x,y)\in X\right\}$ is a rational subset of $\Z^2$.

\begin{defi}
Let $A$ be a nonempty subset of $\{n\in \N; n\geq 2\}$. A sequence $(U(x,y))$ is $A$-automatic if there exists, for each $k\in A$, a $k$-automatic sequence $(V_k(x,y))$, such that $U(x,y)$ is a function of $(V_k(x,y))_k$.
We say that $(U(x,y))$ is $\varnothing$-automatic if it takes values in a finite set $X$ and the preimage of every element of $X$ is a rational subset of $\Z^2$.
\end{defi}

Note that, if $A\subseteq B$, $A$-automaticity implies $B$-automaticity (this includes the case $A=\varnothing$).  

To justify this definition, let us consider a well-known class of cellular automata, called Ledrappier cellular automata in \cite{rowland2020automaticity}. For an integer $m\geq 2$, let us define
\begin{displaymath}
F_m: \left(    \begin{array}{rcl}  
\Z_m^\Z & \to & \Z_m^\Z \\
(r_n)_{n\in \Z} & \mapsto  &  \left( r_n+r_{n-1} \right)_{n\in \Z} 
\end{array}
  \right).
\end{displaymath}

This automaton, starting on the initial configuration $\bar 1$, produces as a spacetime diagram Pascal's triangle modulo $m$, and it has been proven in \cite{takahashi1992self} that, when $m$ is prime, this spacetime diagram is $m$-automatic.  In  \cite{allouche1997automaticity}, it was proven that it is not $m$-automatic when $m$ is not a prime power.  
For instance, Pascal's triangle modulo 6 is not 6-automatic.  However, since, for an integer $n$, $n\pmod 6$ is a function of $n\pmod 2$ and $n\pmod 3$, each cell of Pascal's triangle modulo 6 is a function of the corresponding cells in Pascal's triangles modulo 2 and 3.  Therefore, using the terms of our definition, Pascal's triangle modulo 6 is $\{2,3\}$-automatic.  In fact, it follows from \cite{takahashi1992self} that Pascal's triangle modulo $m$ is $\pi_m$-automatic, where $\pi_m$ is the set of prime divisors of $m$.

A similar statement follows from \cite{Gutschow2010fractal}.  In order to state it, we need to introduce some notation.
For every element $x$ of a finite monoid $(M,\cdot)$, there are least integers $\iindex\geq 0$ and $\period>0$ such that $x^{\iindex+\period}=x^\iindex$;  these are called respectively the threshold and the period of $x$.  We denote $\pi(M)$ the set of prime divisors of periods of elements of $M$.  Note that when $M$ is a group, by Cauchy's theorem, $\pi(M)$ is the set of prime divisors of its order $|M|$.  
We can then state the following corollary of Theorem~\ref{thm:GNW10}.

\begin{prop}\label{prop:corollary}
If $(\Sigma,\cdot)$ is an abelian group, then the double sequence generated by a one-dimensional cellular automaton that is also an endomorphism of $\Sigma^\Z$, starting on a finite initial configuration, is $\pi(\Sigma)$-automatic.
\end{prop}

\begin{proof}
For each prime number $p$, let $\Sigma_p$ be  the Sylow $p$-subgroup of $\Sigma$; then $\Sigma$ is isomorphic to $\prod\limits_p \Sigma_p$, and every endomorphism of $\Sigma^\Z$ factorizes into a product of endomorphisms of its subgroups $\Sigma_p^\Z$. 
\end{proof}

Our first result shows that Proposition~\ref{prop:corollary} remains valid when groups are replaced by commutative monoids.

\begin{thm}\label{thm:main}
If $(\Sigma,\cdot)$ is a commutative monoid, then the double sequence generated by a one-dimensional cellular automaton that is also an endomorphism of $\Sigma^\Z$, starting on a finite initial configuration, is $\pi(\Sigma)$-automatic.
\end{thm}

Theorem~4.5 of \cite{rowland2020automaticity} states that, if $(\Sigma,\cdot)=\Z_p$ and the initial configuration is $p$-automatic for some prime number $p$, then so is the spacetime diagram.  Our second result generalizes this theorem.

\begin{thm}\label{thm:automatic_initial_configuration}
Let $p$ be a prime number and $\Sigma$ a finite commutative monoid  such that $\pi(\Sigma)\subseteq \{p\}$.  Let $F:\Sigma^{\Z}\to \Sigma^{\Z}$ be a cellular automaton that is also an endomorphism of $\Sigma^{\Z}$.  If the initial configuration $c:\Z\to \Sigma$ is $p$-automatic, then so is its spacetime diagram.
\end{thm}

This paper is organized as follows.   Section~\ref{sec:groups} is an erratum of a lemma in \cite{Gutschow2010fractal}, which treated the case of abelian groups:  it can be skipped. Section~\ref{sec:aperiodic} treats the case ``orthogonal" to groups, namely aperiodic monoids.  It is then shown in Section~\ref{sec:free_monoids} how these two base cases can be brought together to treat the case of free commutative $(\iindex,\period)$-monoids.  Section~\ref{sec:general_case} concludes the proof of Theorem~\ref{thm:main} and Section~\ref{sec:automatic_initial} is devoted to the proof of Theorem~\ref{thm:automatic_initial_configuration}.

\section{Groups}\label{sec:groups}

This section is essentially an erratum of a proposition of \cite{Gutschow2010fractal}.  However, the main theorem of~\cite{Gutschow2010fractal}, reformulated as Theorem~\ref{thm:GNW10} of the present paper, stands and its proof remains essentially correct, so the reader may skip to Section~\ref{sec:aperiodic}.

Let us begin by stating and proving the following common grouping property; its use can be traced back at least to \cite{willson1987computing}.

\begin{prop}\label{prop:grouping}
Let $k\geq 2$ and $d\geq 1$ be integers.  Let $E$ be a finite set and $U:\Z^d\to E$. 
Suppose there exists a finite set $I\subseteq \Z^d$ and, for every $\multi{s}\in\intint{0,k-1}^{d}$,  $\epsilon_{\multi{s}}:E^I \to E$ such that
\begin{equation}\label{eq:prop_grouping}
\forall \multi{n}\in \Z^d\quad U(k\multi{n} + \multi{s}) =  \epsilon_\multi{s} (U(\multi{n}-\multi{i})_{i\in I}).
\end{equation}

Then $U$ is $k$-automatic.
\end{prop}

\begin{proof}
Notice that (\ref{eq:prop_grouping}) can be so reformulated:
\begin{equation}
\forall\multi{n}\in\Z^d\quad U(\multi{n})= \epsilon_{\multi{n}\bmod k} \left(U\left(\floor{\dfrac{1}{k}\multi{n}}-\multi{i}\right)_{i\in I}\right).
\end{equation}

For a set $J\subseteq \Z^d$, define $V_J(\multi{n}) = (U(\multi{n}-\multi{j}))_{\multi{j}\in J}$.
For any $\multi{s}\in\intint{0,k-1}^{d}$, 
\begin{align*}
V_J(k\multi{n} + \multi{s}) &=  \left(U(k\multi{n}+\multi{s}-\multi{j})\right)_{\multi{j}\in J} \\
&= \left( \epsilon_{(k\multi{n}+\multi{s}-\multi{j})\bmod k} \left(U\left(\floor{\dfrac{1}{k}(k\multi{n}+\multi{s}-\multi{j})}-\multi{i}\right)_{i\in I}\right)\right)_{\multi{j}\in J} \\
& = \left(\epsilon_{ (\multi{s}-\multi{j})\bmod k } \left(U \left(\multi{n}+\floor{\dfrac{1}{k}(\multi{s}-\multi{j})}-\multi{i}\right)_{\multi{i}\in I}\right)\right)_{\multi{j}\in J} \\
\end{align*}

From this expression we deduce that if $J$ is finite and such that 

\begin{equation} \label{eq:grouping}
\forall \multi{s}\in \intint{0,k-1}^{d}\;\forall \multi{i}\in I\;\forall \multi{j}\in J\quad \multi{i}-\floor{\dfrac{1}{k}(\multi{s}-\multi{j})} \in J
\end{equation}

then $V_J$ is $k$-automatic.  If, moreover, $(0,\ldots,0)\in J$, then $U(\multi{n})$ is a function of $V_J(\multi{n})$, so $U$ is itself $k$-automatic.

Such a set $J$ indeed exists. Let $a\in\N$ be such that $I\subseteq \intint{-a,a}^d$.
Let $b$ be an integer larger than $\dfrac{a+1}{1-\frac{1}{k}}$ and $J=\intint{-b,b}^d$.  Obviously, $J$ is finite and contains $(0,\ldots,0)$.
It also fulfills (\ref{eq:grouping}), because for any $s\in\intint{0,k-1}$, $i\in \intint{-a,a}$ and $j\in\intint{-b,b}$,  we have:

\begin{align*}
-a - \left\lfloor \dfrac{1}{k} ((k-1)-(-b))\right\rfloor  \leq &i -  \left\lfloor \dfrac{1}{k}(s-j)\right\rfloor \leq  a - \left\lfloor \dfrac{1}{k}(0-b)\right\rfloor  \\
-a - \left\lfloor 1 - \dfrac{1}{k} (-b+1) \right\rfloor  \leq &i -  \left\lfloor \dfrac{1}{k}(s-j)\right\rfloor \leq  a + \left\lceil \dfrac{1}{k}b \right\rceil  \\
-a - 1 + \left\lceil \dfrac{1}{k}(-b+1)\right\rceil   \leq &i -  \left\lfloor \dfrac{1}{k}(s-j)\right\rfloor  \leq  a + 1 + \dfrac{1}{k}b \\
-(a +1)+ \dfrac{1}{k}(-b+1)  \leq &i -  \left\lfloor \dfrac{1}{k}(s-j)\right\rfloor \leq b\left(1-\dfrac{1}{k}\right) + \dfrac{1}{k}b \\
-b\left(1-\dfrac{1}{k}\right) + \dfrac{1}{k}(-b+1) \leq &i -  \left\lfloor \dfrac{1}{k}(s-j)\right\rfloor  \leq  b \\
-b  + \dfrac{1}{k} \leq &i -  \left\lfloor \dfrac{1}{k}(s-j)\right\rfloor \leq b \\
-b \leq &i -  \left\lfloor \dfrac{1}{k}(s-j)\right\rfloor \leq b .
\end{align*}

We can therefore conclude that $U$ is $k$-automatic.
\end{proof}

Let us now concentrate on an erroneous statement that can be found in \cite{Gutschow2010fractal}, namely Proposition~4.  What is stated in that paper is the following:

\begin{center}
\fbox{ 
\begin{minipage}[c]{0.8\textwidth}

Let $R$ be a finite commutative ring, $M$ a finite $R$-module, $k$ and $m$ positive integers, $\Lambda$ a finite set of indices, and for $i\in \Lambda$, $f_i:\llbracket m,+\infty\llbracket \to \Z$ and $g_i:\llbracket  m,+\infty \llbracket \to \N$ such that for all $y\in \llbracket m;+\infty \llbracket$ and $t\in\llbracket 0,k-1\rrbracket$,

\begin{itemize}
\item $g_i(y)<y$;
\item $f_i(ky+t) = k f_i(y)$ and $g_i(ky+t)=kg_i(y)+t$.
\end{itemize}

For $(x,y)\in\Z\times\N$, let $\Xi_x^y\in M$ be such that when $y\geq m$,
\begin{displaymath}
\Xi_x^y = \sum\limits_{i\in\Lambda} \mu_i  \Xi _{x+f_i(y)}^{g_i(y)} .
\end{displaymath}

Then there exists a finite set $E$ and a function $e:\Z\times\N\to E$ such that
\begin{itemize}
\item $\Xi_x^y$ is a function of $e(x,y)$;
\item for $s,t\in\llbracket 0,k-1\rrbracket$, $e(kx+s,ky+t)$ is a function of $s$, $t$ and $e(x,y)$.
\end{itemize}
\end{minipage}
}
\end{center}

This is false because it implies that the sequence $(\Xi_x^0)_{x\in\Z}\in M^\Z$ is $k$-automatic, whereas it could be any arbitrary sequence; you could in fact define each $\Xi_x^y$, for $x\in\Z$ and $y<m$, to be any arbitrary element of $M$.  The proposition can be fixed by assuming that the values $\Xi_x^y$, for $y<m$, are almost all equal to 0.  The proof remains essentially the same, and this mistake does not impact other statements of \cite{Gutschow2010fractal}, because the proposition is only applied to cases where the added assumption is true.  Let us seize this opportunity to fix the proposition and generalize it from $R$-modules to commutative monoids.

In this proposition, we will use the additive notation for the monoid operation.

\begin{prop}\label{prop:corrigee}
Let $(M,+,0)$ be a finite commutative monoid, $m$ a positive integer, $k\geq 2$ an integer, $\Lambda$ a finite set of indices, and for $i\in \Lambda$, $f_i:\llbracket m,+\infty\llbracket \to \Z$ and $g_i:\llbracket  m,+\infty \llbracket \to \N$ such that for all $y\in \llbracket m;+\infty \llbracket$ and $t\in\llbracket 0,k-1\rrbracket$,

\begin{itemize}
\item $g_i(y)<y$;
\item $f_i(ky+t) = k f_i(y)$ and $g_i(ky+t)=kg_i(y)+t$.
\end{itemize}

Let $(\varphi_i)_{i\in\Lambda}$ be a family of endomorphisms of $M$, and $\Xi:\Z\times\N \to M$ be such that $\{(x,y)\in \Z\times \llbracket 0 ;m-1\rrbracket ; \Xi(x,y)\neq 0 \}$ is finite and when $y\geq m$,
\begin{equation}
\Xi(x,y) = \sum\limits_{i\in\Lambda} \varphi_i \circ \Xi \left(x+f_i(y),g_i(y) \right) . \label{eq:xi_recursion}
\end{equation}

Then $\Xi$ is $k$-automatic.

\end{prop}

Let us define by induction on $y$, for $j,y\in \N$ and $x\in\Z$, the following endomorphisms $\alpha_{j}(x,y)$ of $M$:
\begin{itemize}
\item if $y<m$ then 
 $\alpha_{j}(x,y) = \left\{  \begin{array}{ll}
\id & \text{if $(0,j)=(x,y)$} \\ 
0 & \text{otherwise}
\end{array}\right.,$

\item if $y\geq m$ then $\alpha_{j}(x,y) =  \sum\limits_{i\in\Lambda} \varphi_i \circ \alpha_j(x+f_i(y),g_i(y)) $
\end{itemize}

In some sense, the proof we present here is more direct than the one proposed in \cite{Gutschow2010fractal}; however, that article explains in some detail the intuition behind the definition of these $\alpha_j$ endomorphisms, so that the reader finding such explanations lacking in the present paper can be redirected to \cite{Gutschow2010fractal}.

The plan of the proof is simple: we prove three lemmas, then conclude by showing that Proposition~\ref{prop:corrigee} results from these three lemmas together with Proposition~\ref{prop:grouping}.

\begin{lemma}\label{lemma1}
For every $y\in\N$, almost all the values $\alpha_j(x,y)$, for $j<m$ and $x\in\Z$, are equal to 0.
\end{lemma}
\begin{proof}
Let us prove it by induction on $y$.
\begin{itemize}
\item  If $y<m$, then there is at most one nonzero $\alpha_j(x,y)$ for $j<m$ and $x\in\Z$, namely $\alpha_y(0,y)=\id$.

\item Let $y\geq m$ and suppose that for every $y'<y$, almost all the values $\alpha_j(x,y')$, for $j<m$ and $x\in\Z$, are equal to 0.

Then there exists $B$ such that whenever $y'<y$, $j<m$ and $|x|>B$, $\alpha_j(x,y')=0$.
Likewise, $\Lambda$ being a finite set, there exists $C$ such that for every $i\in\Lambda$, $|f_i(y)|\leq C$.

Let $x$ be such that $|x|>B+C$. Then, for every $j<m$, $|x+f_i(y)|>B$.  Thus, for every $i\in\Lambda$,  since $g_i(y)<y$, $\alpha_j(x+f_i(y),g_i(y))=0$. Therefore
$\alpha_{j}(x,y) =  \sum\limits_{i\in\Lambda} \varphi_i \circ \alpha_j(x+f_i(y),g_i(y)) = 0$. This proves that almost all the values $\alpha_j(x,y)$, for $j<m$ and $x\in\Z$, are equal to 0, concluding the induction step and the proof of Lemma~\ref{lemma1}.
\end{itemize}
\end{proof}

\begin{lemma}\label{lemma2}
 For every $(x,y)\in\Z\times \N$,  $\Xi(x,y) = \sum\limits_i  \sum\limits_{j=0}^{m-1} \alpha_{j}(x-i,y)\circ  \Xi(i,j)$
\end{lemma}

\begin{proof}
Let $\mathcal{P}(y)$ be the assertion ``for every $x\in\Z$, $\Xi(x,y) = \sum\limits_i  \sum\limits_{j=0}^{m-1} \alpha_{j}(x-i,y)\circ  \Xi(i,j)$".
Let us prove it  by induction on $y$. 

\begin{itemize}
\item If $y< m$, then 

\begin{displaymath}
\sum\limits_i  \sum\limits_{j=0}^{m-1} \alpha_{j}(x-i,y)\circ  \Xi(i,j) = \sum\limits_i  \sum\limits_{j=0}^{m-1}   \delta_{(0,j)(x-i,y)} \Xi(i,j),
\end{displaymath}
where $\delta$ is the Kronecker delta, from which we get $\sum\limits_i  \sum\limits_{j=0}^{m-1} \alpha_{j}(x-i,y)\circ  \Xi(i,j) = \Xi(x,y)$. 

So $\mathcal{P}(y)$ is true.

\item Let $y\geq m$ and suppose $\mathcal{P}(y')$ is true for every $y'<y$.  Let $x\in\Z$.
By (\ref{eq:xi_recursion}), we have  
\begin{displaymath}
\Xi(x,y)  = \sum\limits_{i\in\Lambda} \varphi_i \circ \Xi \left(x+f_i(y),g_i(y) \right)
\end{displaymath}

Since $g_i(y)<y$, we can apply our induction hypothesis to get 
\begin{align*}
\Xi(x,y)& = \sum\limits_{i\in\Lambda} \varphi_i \circ \left( \sum\limits_{i'}  \sum\limits_{j=0}^{m-1} \alpha_{j}(x+f_i(y)-i',g_i(y))\circ  \Xi(i',j)  \right) \\
& =  \sum\limits_{i\in\Lambda}   \sum\limits_{i'}  \sum\limits_{j=0}^{m-1}   \varphi_i \circ \alpha_{j}(x+f_i(y)-i',g_i(y))\circ  \Xi(i',j)   \\
& =    \sum\limits_{i'}  \sum\limits_{j=0}^{m-1}   \left( \sum\limits_{i\in\Lambda}  \varphi_i \circ \alpha_{j}((x-i')+f_i(y),g_i(y))\right)\circ  \Xi(i',j) \\
\Xi(x,y) & =  \sum\limits_{i'}  \sum\limits_{j=0}^{m-1}  \alpha_j(x-i',y) \circ  \Xi(i',j) \\
\end{align*}

Therefore $\mathcal{P}(y)$ is true. This completes the induction step and hence, the proof of Lemma~\ref{lemma2}.
\end{itemize}
\end{proof}

\begin{lemma}\label{lemma3}
For every $x\in\Z$, $y\in \N$, $j<m$ and $s,t\in\intint{0,k-1}$,
\[ \alpha_{j}(kx+s,ky+t) =  \sum\limits_{i'} \sum\limits_{j'=0}^{m-1}\alpha_{j'}(x-i',y)  \circ \alpha_{j}(ki'+s,kj'+t).\]
\end{lemma}

\begin{proof}
Let us prove it by induction on $y$.  Let $\mathcal{P}(y)$ be the assertion ``for every $x\in\Z$, $j<m$ and $s,t\in\intint{0,k-1}$, $\alpha_{j}(kx+s,ky+t) =  \sum\limits_{i'} \sum\limits_{j'=0}^{m-1}\alpha_{j'}(x-i',y)  \circ \alpha_{j}(ki'+s,kj'+t)$".

\begin{itemize}
\item If $y<m$, then
$\alpha_{j'}(x-i',y)= \left\{  \begin{array}{ll}
\id & \text{if $(0,j')=(x-i',y)$} \\ 
0 & \text{otherwise}
\end{array}\right.$,  therefore   
\begin{displaymath}
\sum\limits_{i'} \sum\limits_{j'=0}^{m-1}\alpha_{j'}(x-i',y) \circ \alpha_{j}(ki'+s,kj'+t) = \alpha_j(kx+s,ky+t).
\end{displaymath}
So $\mathcal{P}(y)$ is true.

\item Let $y\geq m$ and suppose $\mathcal{P}(y')$ is true for every $y'<y$.  We then have
\begin{align*}
& \alpha_j(kx+s,ky+t) \\
 = &\sum\limits_{i\in\Lambda} \varphi_i \circ \alpha_j(kx+s+f_i(ky+t),g_i(ky+t)) \\
 = & \sum\limits_{i\in\Lambda} \varphi_i \circ \alpha_j(k(x+f_i(y))+s, k g_i(y)+t) \\
 = & \sum\limits_{i\in\Lambda}\varphi_i \circ  \left(\sum\limits_{i'}\sum\limits_{j'=0}^{m-1}\alpha_{j'}(x+f_i(y)-i', g_i(y))\circ \alpha_j(ki'+s,kj'+t)\right) \\
= & \sum\limits_{i'}\sum\limits_{j'=0}^{m-1}\left(  \sum\limits_{i\in\Lambda} \varphi_i \circ \alpha_{j'}(x+f_i(y)-i', g_i(y))\right)\circ \alpha_j(ki'+s,kj'+t) \\
=&\sum\limits_{i'}\sum\limits_{j'=0}^{m-1}\alpha_{j'}(x-i',y)\circ \alpha_j(ki'+s,kj'+t) \\
\end{align*}

So $\mathcal{P}(y)$ is true, which completes the induction step and the proof of Lemma~\ref{lemma3}.
\end{itemize}
\end{proof}

Let $I$ be the finite set $\left\{i\in\Z ; \exists j<m\quad \Xi(i,j)\neq 0\right\}$.    Since, according to Lemma~\ref{lemma1},  almost all the values $\alpha_{j}(ki'+s,kj'+t)$ for $j,j'<m$, $s,t<k$ and $i'\in \Z$, are equal to $0$, there must exist, according to Lemma~\ref{lemma3}, a finite set of indices $I'$ such that, for each $s,t\in \llbracket 0; k-1 \rrbracket$, $(\alpha_j(kx+s-i,ky+t))_{i\in I, j<m}$  is a function of $(\alpha_j(x-i-i',y))_{i\in I, i'\in I',  j< m}$.

The function $U:(x,y)\mapsto (\alpha_j(x-i,y))_{i\in I, j<m}$ can be extended to $\Z^2$ by defining $U(x,y)=\bot$ when $y<0$: $U(kx+s,ky+t)$ is then a function of $s$, $t$ and $(U(x-i',y))_{i'\in I'}$.

According to Proposition~\ref{prop:grouping}, $U$ is therefore $k$-automatic.  And since, according to Lemma~\ref{lemma2}, $\Xi(x,y)$ is a function of $U(x,y)$,  it follows that $\Xi$ is $k$-automatic.

\section{Aperiodic Monoids}\label{sec:aperiodic}

\subsection{First Example}

When looking for a generalization of Theorem~\ref{thm:GNW10}, one has to think of finite commutative monoids that are both easy to understand and quite different from groups. For any positive integer $n$,  let $O_ n$ be the commutative monoid $\left(\intint{0,n-1}, \cdot ,0\right)$, where $a\cdot b = \max(a,b)$.  For every $a\in O_n$, $a\cdot a = a$, so $O_n$ contains no nontrivial subgroup:  In this sense, it is as far as could be from being a group.  Since every element of $O_n$ has period 1, we have $\pi(O_n)=\varnothing$, so in this case, Theorem~\ref{thm:main} states that the double sequence generated by a cellular automaton that is also an endomorphism of $O_n^{\Z}$, starting on a finite initial configuration, is $\varnothing$-automatic.

The endomorphisms of $O_n$ are the nondecreasing functions $f:\intint{0,n-1} \to \intint{0,n-1}$ such that $f(0)=0$.  Let us define the following endomorphisms $f_{0,1}$ of $O_3$:

\begin{center}
\begin{tabular}{|c|c|c|}
\hline
$x$  & $f_0(x)$ & $f_1(x)$ \\
\hline
0 & 0 & 0 \\
1 & 2 & 0 \\
2 & 2 & 1 \\
\hline
\end{tabular}
\end{center}

Together, by Equation~(\ref{eq:globaltransition}), they define a global transition function $F$ that is a cellular automaton and an endomorphism of $O_3^{\Z}$.   Let us run this cellular automaton on the initial configuration $\bar 2\in O_3^{\Z}$ defined by $\bar 2_n = \left\{  \begin{array}{ll}
2 & \text{if $n=0$} \\ 
0 & \text{otherwise}
\end{array}\right.$.

\begin{figure}[htbp]
\centering
\begin{tikzpicture}  [help lines/.style={draw=black},  every node/.style={help lines,rectangle,minimum size=3mm},  cellular automaton/.style={draw=none,row sep=0mm,column sep=0mm},  rst/.style={fill=white,help lines},  exc/.style={fill=blue!70,help lines},  ref/.style={fill=blue!30!white,help lines}]
    \matrix[cellular automaton] {
\node[rst] {2};  \\
\node[rst] {2}; &\node[rst] {1};  \\
\node[rst] {2};  &\node[rst] {2}; \\
\node[rst] {2};  &\node[rst] {2}; & \node[rst] {1};  \\
\node[rst] {2};  &\node[rst] {2}; & \node[rst] {2};  \\
\node[rst] {2};  &\node[rst] {2};  &\node[rst] {2}; & \node[rst] {1};  \\
\node[rst] {2};  &\node[rst] {2};  &\node[rst] {2}; & \node[rst] {2};  \\
\node[rst] {2};  &\node[rst] {2};  &\node[rst] {2};  &\node[rst] {2}; & \node[rst] {1};  \\
\node[rst] {2};  &\node[rst] {2};  &\node[rst] {2};  &\node[rst] {2}; & \node[rst] {2};  \\
\node[rst] {2};  &\node[rst] {2};  &\node[rst] {2};  &\node[rst] {2}; & \node[rst] {2};   & \node[rst] {1};\\
\node[rst] {2};  &\node[rst] {2};  &\node[rst] {2};  &\node[rst] {2}; & \node[rst] {2};   & \node[rst] {2};  \\
    };
  \end{tikzpicture}
\caption{Ten iterations of $F$ on the initial configuration $\bar 2$.  The top cell has coordinates $(0,0)$; time flows downwards.  The neutral element 0 is not depicted.}
\end{figure}

Let $U:\Z\times\N \to O_3$ be the double sequence defined by $U(i,j) = F^j\left(\bar 2\right)_i$.   For $x\in O_3$, let $X_x=U^{-1}(x) =\left\{(i,j)\in \Z\times \N  ;  F^j\left(\bar 2\right)_i = x\right\}$.  Theorem~\ref{thm:main} states that each $X_x$ is a rational subset of $\Z^2$.  In this example, the pattern is quite simple, and this statement is readily seen to be true;  let us see how we can prove it in a generalizable way.

Let us examine $U(1,2)$.  Why is it equal to 2? By definition,  
\begin{equation}
\begin{split}
U(1,2) & = \max \left( f_0\left( U(1,1)\right),  f_1\left(U(0,1)\right )  \right) \\
U(1,1) &= \max \left( f_0\left( U(1,0)\right),  f_1\left(U(0,0)\right )  \right) \\
U(0,1) &=  \max \left( f_0\left( U(0,0)\right),  f_1\left(U(-1,0)\right )  \right)
\end{split}
\end{equation}

Since $f_0$ and $f_1$ are endomorphisms of $O_3$, we can factorize the monoid operation $\max$ and write

\begin{equation}
U(1,2) = \max \left( f_0f_0\left( U(1,0)\right), f_0f_1\left(U(0,0)\right ) , f_1f_0\left(U(0,0)\right ),  f_1f_1\left(U(-1,0)\right )  \right)
\end{equation}

By induction, we thus get that, for any $(i,j)\in\Z\times\N$,
\begin{equation}
U(i,j) = \max\left\{  f_{x_j}f_{x_{j-1}}\cdots f_{x_1}\left(U\left(i-\sum\limits_{k=1}^j x_k,0\right)\right)  ;  x_1,\ldots,x_j \in \{0,1\} \right\}
\end{equation}

Since the initial configuration is $\bar 2$, $U(i,0)= \left\{  \begin{array}{ll}
2 & \text{if $i=0$} \\ 
0 & \text{otherwise}
\end{array}\right.$, so we get

\begin{equation}\label{eq:U(i,j)}
U(i,j) = \max\left\{  f_{x_j}f_{x_{j-1}}\cdots f_{x_1}(2) ;\sum\limits_{k=1}^j x_k= i \right\}.
\end{equation}

With this formula in mind, computing $U(i,j)$ is like playing the following game.  You start from cell $(0,0)$, and have to end up in cell $(i,j)$.  You have two moves at your disposal: you can either go down one cell $\downarrow$ , or down and right $\searrow$.  You also start with the number 2; each time you go $\downarrow$, you have to apply $f_0$ to your number; each time you go $\searrow$, you have to apply $f_1$.  The goal of the game is to get the maximum number at the end.  What happens to the number you are holding during the course of this game can be described by the automaton depicted in Fig.~\ref{fig:automate1}.

\begin{figure}[htbp]
\centering
\begin{tikzpicture}[->,>=stealth,node distance = 3cm,initial text=$ $]
\node[state, initial] (q2) {$2$};
\node[state, right of=q2] (q1) {$1$};
\node[state, right of=q1] (q0) {$0$};
\draw (q2) edge[loop above] node{$\downarrow$} (q2)
(q2) edge[bend left, above] node{$\searrow$} (q1)
(q1) edge[bend left, below] node{$\downarrow$} (q2)
(q1) edge[above] node{$\searrow$} (q0)
(q0) edge[loop above] node{$\downarrow,\searrow$} (q0);
\end{tikzpicture}
\caption{Finite automaton describing the paths of computation.\label{fig:automate1}}
\end{figure}

According to Equation~(\ref{eq:U(i,j)}), $U(i,j)$ is thus the maximum number that you can reach by reading a word on $\{\downarrow,\searrow\}$ that describes a path from cell $(0,0)$ to cell $(i,j)$, i.e., an anagram of $\searrow^i\downarrow^{j-i}$.

For each $x\in O_3$, let $\mathcal{L}_x$ be the rational set of words over $\{\downarrow,\searrow\}$ whose output by this finite automaton is $x$: $\mathcal{L}_2 = (\downarrow|\searrow\downarrow)^*$ , $\mathcal{L}_1  = (\downarrow |\searrow \downarrow )^*\searrow $,  $\mathcal{L}_0  = (\downarrow |\searrow \downarrow )^*\searrow \searrow (\downarrow |\searrow )^*$.    Now consider the monoid morphism $\varphi: \{\downarrow,\searrow\}^* \to \Z\times\N$ defined by $\varphi(\downarrow )=(0,1)$ and $\varphi(\searrow)=(1,1)$.  As direct images of rational sets under a monoid morphism, $\varphi(\mathcal{L}_0)$,  $\varphi(\mathcal{L}_1)$ and  $\varphi(\mathcal{L}_2)$ are rational subsets of $\Z\times \N$.  Figure~\ref{fig:troisphi} represents these three sets.

\begin{figure}[htbp]
\centering
\begin{tikzpicture}
[help lines/.style={draw=black},  every node/.style={help lines,rectangle,minimum size=8mm},  cellular automaton/.style={draw=none,row sep=0mm,column sep=0mm},  r/.style={fill=white,help lines}]
    \matrix[cellular automaton] {
\node[r] {2};  \\
\node[r] {2}; & \node[r] {1}; \\
\node[r] {2};  &\node[r] {12}; & \node[r] {0} ; \\
\node[r] {2};  &\node[r] {12}; & \node[r] {01}; & \node[r] {0}; \\
\node[r] {2};  &\node[r] {12}; & \node[r] {012}; & \node[r] {0} ; & \node[r] {0}; \\
\node[r] {2};  &\node[r] {12};  &\node[r] {012}; & \node[r] {01};  & \node[r] {0};& \node[r] {0};\\
\node[r] {2};  &\node[r] {12};  &\node[r] {012}; & \node[r] {012};& \node[r] {0};& \node[r] {0};& \node[r] {0};  \\
\node[r] {2};  &\node[r] {12};  &\node[r] {012};  &\node[r] {012};& \node[r] {01} ; & \node[r] {0};& \node[r] {0};& \node[r] {0};\\
\node[r] {2};  &\node[r] {12};  &\node[r] {012};  &\node[r] {012}; & \node[r] {012};& \node[r] {0};& \node[r] {0};& \node[r] {0};& \node[r] {0};  \\
};

\end{tikzpicture}
\caption{For $0\leq i\leq j$, the cell $(i,j)$ contains $x$ iff $(i,j)\in\varphi(\mathcal{L}_x)$.  The top cell has coordinates $(0,0)$.\label{fig:troisphi}}
\end{figure}

$X_2 = \varphi(\mathcal{L}_2)$ is therefore rational; and so are $X_1= \varphi(\mathcal{L}_1) \setminus \varphi(\mathcal{L}_2)$ and $X_0 = (\Z\times\N) \setminus \left( X_1\cup  X_2\right)$ because boolean combinations of rational subsets of $\Z^2$ are rational.

\subsection{Second Example}\label{sec:deuxieme_exemple}

In our first example $O_3$, Equation~(\ref{eq:U(i,j)}) has the crucial property that it expresses $U(i,j)$ as a function of the set of the states that are attainable in a given finite automaton.  This is possible because, for every $a\in O_n$, $a\cdot a=a$.   We will however need to treat more general cases where this condition is not fulfilled.  Let us consider, once again, a very basic example, that will illustrate the problem and its solution.  For any positive integer $n$, let $P_n =  \left<a| a^{n-1}=a^n \right>$.  In order to alleviate notation, we can discard $a$ and write only its exponent. In this manner, $P_n$ is the commutative monoid $\left(\intint{0,n-1}, \cdot ,0\right)$, where $x\cdot y = \min(x+y,n-1)$.  Like $O_n$, $P_n$ has no nontrivial subgroup, and is as such as far as could be from being a group.  Every element of $P_n$ has period 1, so Theorem~\ref{thm:main} states that the double sequence generated by a cellular automaton that is also an endomorphism of $P_n^{\Z}$, starting on a finite initial configuration, is $\varnothing$-automatic.  Since $1$ is a generator of $P_n$, the endomorphisms of $P_n$ are defined by their image of $1$.

Let us work in $P_3$, and define its endomorphisms $g_0=g_1=\id_{P_3}$.  They define a global transition function $G$ that is a cellular automaton and an endomorphism of $P_3^{\Z}$.   When we run this cellular automaton on the initial configuration $\bar 1$,  we get Pascal's triangle capped at 2.

\begin{figure}[htbp]
\centering
\begin{tikzpicture}  [help lines/.style={draw=black},  every node/.style={help lines,rectangle,minimum size=3mm},  cellular automaton/.style={draw=none,row sep=0mm,column sep=0mm},  rst/.style={fill=white,help lines},  exc/.style={fill=blue!70,help lines},  ref/.style={fill=blue!30!white,help lines}]
    \matrix[cellular automaton] {
\node[rst] {1};  \\
\node[rst] {1}; &\node[rst] {1};  \\
\node[rst] {1};  &\node[rst] {2}; ; &\node[rst] {1};  \\
\node[rst] {1};  &\node[rst] {2}; & \node[rst] {2};  ; &\node[rst] {1}; \\
\node[rst] {1};  &\node[rst] {2}; & \node[rst] {2};   & \node[rst] {2};  &\node[rst] {1}; \\
\node[rst] {1};  &\node[rst] {2};  &\node[rst] {2}; & \node[rst] {2};  & \node[rst] {2};  &\node[rst] {1};  \\
\node[rst] {1};  &\node[rst] {2};  &\node[rst] {2}; & \node[rst] {2};  & \node[rst] {2};  & \node[rst] {2};  &\node[rst] {1}; \\
\node[rst] {1};  &\node[rst] {2};  &\node[rst] {2};  &\node[rst] {2}; & \node[rst] {2}; & \node[rst] {2};  & \node[rst] {2};  &\node[rst] {1}; \\
\node[rst] {1};  &\node[rst] {2};  &\node[rst] {2};  &\node[rst] {2}; & \node[rst] {2}; & \node[rst] {2}; & \node[rst] {2};  & \node[rst] {2};  &\node[rst] {1}; \\
\node[rst] {1};  &\node[rst] {2};  &\node[rst] {2};  &\node[rst] {2}; & \node[rst] {2};   & \node[rst] {2};& \node[rst] {2}; & \node[rst] {2};  & \node[rst] {2};  &\node[rst] {1};\\
\node[rst] {1};  &\node[rst] {2};  &\node[rst] {2};  &\node[rst] {2}; & \node[rst] {2};   & \node[rst] {2};  & \node[rst] {2};& \node[rst] {2}; & \node[rst] {2};  & \node[rst] {2};  &\node[rst] {1}; \\
    };
  \end{tikzpicture}
\caption{Ten iterations of $G$ on the initial configuration $\bar 1$.  The top cell has coordinates $(0,0)$; time flows downwards.  The neutral element $0$ is not depicted.}
\end{figure}

If we follow the same logic as in the previous example, we get the following finite automaton:

\begin{center}
\centering
\begin{tikzpicture}[->,>=stealth,node distance = 3cm,initial text=$ $]
\node[state, initial] (q2) {1};
\draw (q2) edge[loop above] node{$\downarrow,\searrow$} (q2);
\end{tikzpicture}
\end{center}

It is clearly irrelevant in this case.  For $(i,j)\in\Z\times \N$, let $V(i,j)=G^j(\bar 1)_i$.   Instead of Equation~\ref{eq:U(i,j)}, we get 

\begin{equation}\label{eq:V(i,j)}
V(i,j) = \min\left( 2, \sum\limits_{x_1+\ldots+x_j= i}  g_{x_j}g_{x_{j-1}}\cdots g_{x_1}(1) \right).
\end{equation}

For $x\in P_3$, let $X_x=V^{-1}(x) =\left\{(i,j)\in \Z\times \N  ;  G^j\left(\bar 1\right)_i = x\right\}$.  Theorem~\ref{thm:main} states that each $X_x$ is a rational subset of $\Z^2$.  Again, this is obviously true in this example, but let us see how we will prove it in the general case.

If we describe the same game as in the previous section for computing $V(i,j)$, the number that the player holds,  $ g_{x_j}g_{x_{j-1}}\cdots g_{x_1}(1)$, is always $1$, since $g_0(1)=g_1(1)=1$.  Now, the question is whether the cell $(i,j)$ can be reached by a unique path or by at least two paths.  We therefore need a finite automaton that keeps track not only of single paths of computations but of pairs of paths of computation, so as to be able to tell if the state 1 is reachable at least twice.  We now have a finite automaton on the alphabet $\Sigma = \left\{(\downarrow,\downarrow), (\searrow,\searrow), (\downarrow,\searrow), (\searrow,\downarrow)\right\}$, whose states not only contain the information about  $g_{x_j}g_{x_{j-1}}\cdots g_{x_1}(1)$ and $g_{y_j}g_{y_{j-1}}\cdots g_{y_1}(1)$ (which are anyway both always equal to 1 in our example), but also about whether both paths are identical to each other, i.e., whether for every $k\in\{1,\ldots,j\}$, $x_k=y_k$.  We get the automaton depicted in Figure~\ref{fig:example2}.

\begin{figure}[htbp]
\centering
\begin{tikzpicture}[->,>=stealth,node distance = 3cm,initial text=$ $]
\node[state, initial] (q2) {=};
\node[state, right of =q2] (q1) {$\neq$};
\draw (q2) edge[loop above] node{$(\downarrow,\downarrow)$, $(\searrow,\searrow)$} (q2) ;
\draw (q2) edge[bend left, above] node{$(\downarrow,\searrow)$, $(\searrow,\downarrow)$} (q1) ;
\draw (q1) edge[loop above] node{$(\downarrow,\downarrow)$, $(\searrow,\searrow)$} (q1) ;
\draw (q1) edge[loop below] node{$(\downarrow,\searrow)$, $(\searrow,\downarrow)$} (q1) ;
\end{tikzpicture}
\caption{Finite automaton describing pairs of paths\label{fig:example2}}
\end{figure}

The state $=$, which is the case where the pair of paths is identical, evaluates to 1, whereas $\neq$ evaluates to 2.   The languages recognized by this automaton are  $\mathcal{L}_{=} = \left((\downarrow,\downarrow)| (\searrow,\searrow)\right)^*$ and $\mathcal{L}_{\neq} = \Sigma^* \left((\downarrow,\searrow)| (\searrow,\downarrow)\right) \Sigma^*$.  We define the monoid morphism $\varphi: \Sigma^* \to \Z^2\times\N$ by 

\begin{displaymath}
\varphi:\begin{array}{rcl}
(\downarrow,\downarrow) &\mapsto & (0,0,1) \\
(\searrow,\searrow) &\mapsto & (1,1,1)\\
(\downarrow,\searrow) &\mapsto &(0,1,1) \\
(\searrow,\downarrow) & \mapsto & (1,0,1) \\
\end{array}.
\end{displaymath}

$\varphi(\mathcal{L}_{=})$ and $\varphi(\mathcal{L}_{\neq})$ are rational subsets of $\Z^3$.  For  $s\in\{=,\neq\}$, let 

\begin{equation}
\Delta_{s}=\left\{ (i,j) \in\Z\times\N; (i,i,j)\in \varphi(\mathcal{L}_{s})\right\}.
\end{equation}

$\Delta_{=}$ and $\Delta_{\neq}$ are rational subsets of $\Z^2$.  By definition, $(i,j)\in \Delta_{=}$ iff there is at least one path from $(0,0)$ to $(i,j)$, and $(i,j)\in \Delta_{\neq}$ iff there are at least two different paths from $(0,0)$ to $(i,j)$.   We therefore have $X_2 = \Delta_{\neq}$ and $X_1 = \Delta_{=}\setminus \Delta_{\neq}$, which proves they are rational subsets of $\Z^2$.  Since $X_0=(\Z\times \N)\setminus (X_1\cup X_2)$, it is also rational, which concludes the proof.

\subsection{General Aperiodic Case}

A monoid $M$ is aperiodic if the period of all of its elements is $1$: for every $a\in M$, there exists $n>0$ such that $a^{n+1}=a^n$; when $M$ is finite, this is equivalent to saying that $M$ has no nontrivial subgroup.
On any commutative monoid $M$, one can define a preorder: $x\leq y$ iff there exists $z\in M$ such that $x=yz$.  Let $1$ be the identity element of $M$: for every $x\in M$, $x\leq 1$.

Suppose $M$ is a commutative aperiodic monoid.  Let $a,b\in M$ be such that $a\leq b\leq a$.  Let $x,y\in M$ be such that $a=bx$ and $b=ay$.
Then $a=a(xy)=a(xy)^n$ for every $n\geq 0$.  Let $n$ be such that $y^{n+1}=y^n$.   Then $a=a(xy)^{n}=ax^ny^n=ax^ny^{n+1}=a(xy)^ny=ay=b$.  Therefore $\leq$ is a partial order on $M$.  In the following proposition, ``$\min$" refers to this partial order. Note that $\min\varnothing = \max M = 1$.

\begin{prop}
Let $M$ be a finite commutative aperiodic monoid.  Then there exists $\omega\in\N$ such that, for every $n\in\N$ and every finite sequence $(x_i)\in M^n$,
\begin{equation}\label{eq:minimumomega}
\prod\limits_{i=1}^n x_i = \min \left\{\prod\limits_{i\in A} x_i ; A\subseteq \intint{1,n} , |A|\leq \omega \right\}.
\end{equation}
\end{prop}

\begin{proof}
Let us rewrite $\prod\limits_{i=1}^n x_i= \prod\limits_{x\in M} x^{\alpha_x}$, where $\alpha_x$ is the number of occurrences of $x$ in $(x_i)_{i=1\ldots n}$.   Let $N>0$ be such that, for every $x\in M$, $x^{N+1}=x^N$.   Then  $\prod\limits_{i=1}^n x_i=\prod\limits_{x ; \alpha_x > 0}  x^{\min(\alpha_x,N)}$.  The proposition is therefore true for $\omega=N\times |M|$. 
\end{proof}

The upper bound $\omega\leq N\times |M|$  is very crude, but as long as we do not care for efficiency, it will do.  

\begin{prop}\label{prop:aperiodic}
Let $\Sigma$ be a finite commutative aperiodic monoid, $I$ a finite subset of $\Z$ and $(f_i)_{i\in I}$ a family of endomorphisms of $\Sigma$.
Let $F:\Sigma^\Z\to \Sigma^\Z$ be the cellular automaton defined by

\begin{displaymath} 
F(r)_n=  \prod\limits_{i\in I}  f_i(r_{n-i})
\end{displaymath}

Then, on any finite initial configuration, the spacetime diagram generated by $F$ is $\varnothing$-automatic.
\end{prop}

\begin{proof}
Let $\omega$ be such that Equation~(\ref{eq:minimumomega}) holds in $\Sigma$.
Let $c$ be an initial configuration with finite support.  For every $(i,j)\in\Z\times\N$,

\begin{equation}\label{eq:proof_aperiodic}
F^j(c)_i = \prod\limits_{\substack{x_0\in \supp(c) \\  x_1,\ldots,x_j  \in I \\ x_0+\cdots+x_j = i}}  f_{x_j}f_{x_{j-1}}\cdots f_{x_1}(c_{x_0}).
\end{equation}


Now, let us define a deterministic finite automaton with output with the following characteristics:
\begin{itemize}
\item Its set of states is $\{q_0\} \sqcup \Sigma^\omega\times \mathcal{P}\left( \intint{1,\omega}^{[2]}\right)$, where $X^{[2]}$ denotes the set of unordered pairs of $X$ and $\mathcal{P}(X)$ the power set of $X$.
\item Its alphabet is the disjoint union $(\supp(c))^{\omega}\sqcup I^{\omega}$.
\item Its transition function is defined in the following way: 
\begin{itemize}
\item for each $(x_1,\ldots,x_\omega)\in \supp(c)^\omega$, 

\begin{displaymath}
\delta(q_0,(x_1,\ldots,x_\omega))=\left( (c_{x_1},\ldots,c_{x_\omega}) ,   \left\{ \{i,j\} ; x_i \neq x_j  \right\}\right);
\end{displaymath}
\item for each $(a_1,\ldots,a_\omega)\in \Sigma^\omega$, $A\subseteq  \intint{1,\omega}^{[2]}$ and $(x_1,\ldots,x_\omega)\in I^\omega$,

\begin{align*}
& \delta\left(\left( (a_{1},\ldots,a_{\omega}) ,  A \right),(x_1,\ldots,x_\omega)\right) \\
= & \left( (f_{x_1}(a_1),\ldots,f_{x_\omega}(a_\omega)) ,   A\cup \left\{ \{i,j\} ; x_i \neq x_j  \right\}\right); \\
\end{align*}
\item in other cases, $\delta$ is undefined.
\end{itemize}

\item Its output set is $\Sigma$. The output of state $((a_1,\ldots,a_\omega),A)$ is $\prod\limits_{i\in B} a_i$, where $B$ is the lexicographically least maximal subset of $\intint{1,\omega}$ such that $B^{[2]}\subseteq A$.
\end{itemize}

The idea is that this automaton, instead of following one ``path of computation" of the cellular automaton, follows $\omega$ at once, and keeps track of which pairs of branches are distinct --- that is the role of the set of unordered pairs.  The output of a state is then obtained by choosing a maximal subset of pairwise distinct paths, and multiplying their outputs.

Let $M=\left(\supp(c)^\omega\sqcup (I^\omega)\right)^*$ be the Kleene star of the alphabet of our automaton, and for $a\in\Sigma$, let $\mathcal{L}_a\subseteq M$ be the rational set of words whose output is $a$.
Let $w\in M$ be a word of positive length $j+1$.   One can write $w=(x^1_0,\ldots,x^\omega_0)(x^1_1,\ldots,x^\omega_1)\ldots (x^1_j,\ldots,x^\omega_j)$.    Let $B$ be a maximal subset of $\intint{1;\omega}$ such that the $(j+1)$-tuples $(x^k_0,\ldots,x^k_j)$, for $k\in B$, are pairwise distinct.   By construction, the output of the automaton upon reading $w$ is then $\prod \limits_{k\in B}  f_{x^k_j}f_{x^k_{j-1}}\cdots f_{x^k_1}(c_{x^k_0})$.


Each one of these tuples corresponds to one path of computation, but the product is meaningless unless these paths end up in the same cell of the spacetime diagram.
In order to make sure they do, let us consider the monoid morphism $\varphi: M \to \Z^\omega\times\N$ defined by 

\begin{equation}
\varphi(\bar{x})=\left\{ \begin{array}{ll}
(\bar{x},0) & \text{if $\bar x\in \supp(c)^\omega$}\\
(\bar x,1) & \text{ if $\bar x \in I^\omega$} \\
\end{array}\right..
\end{equation}
\
For $a\in\Sigma$, let $\Delta_a\subseteq\Z\times\N$ be the diagonal of $\varphi(\mathcal{L}_a)$, i.e.,

\begin{equation}
(i,j)\in\Delta_a\iff (i,\ldots,i,j)\in \varphi(\mathcal{L}_a).
\end{equation}
Since $\mathcal{L}_a$ is rational, so is its image under a morphism, namely $\varphi(\mathcal{L}_a)$, and thus, by Proposition~\ref{prop:rational}, so is $\Delta_a$.

Let $(i,j)\in\Z\times\N$ and $X^j_i=\left\{(x_0,\ldots,x_j)\in \supp(c)\times I^j ; x_0+\ldots + x_j = i\right\}$.  
$(i,j)\in \Delta_a$ if and only if, for some $n\in \intint{1,\omega}$, there are $n$ distinct tuples $(x^k_0,\ldots,x^k_j)\in X^j_i$, for $k\in\intint{1,n}$, such that $a=\prod\limits_{k=1}^n   f_{x^k_j}f_{x^k_{j-1}}\cdots f_{x^k_1}(c_{x^k_0})$.
In other words, $(i,j)\in\Delta_a$ iff there exists $A\subseteq X^j_i$ such that $0<|A|\leq \omega$ and $a=\prod\limits_{(x_0,\ldots,x_j)\in A}   f_{x_j}f_{x_{j-1}}\cdots f_{x_1}(c_{x_0})$.

On the other hand, from (\ref{eq:proof_aperiodic}) and  (\ref{eq:minimumomega}) we get 

\begin{equation}
F^j(c)_i = \min \left\{\prod\limits_{(x_0,\ldots,x_j)\in A}  f_{x_j}f_{x_{j-1}}\cdots f_{x_1}(c_{x_0})  ; A\subseteq X^j_i , |A|\leq \omega \right\}.
\end{equation}

We thus get $F^j(c)_i = \min\left\{  a \in \Sigma ; (i,j)\in \Delta_a\right\}$.  Since $\Delta_a$ is a rational subset of $\Z\times\N$ for every $a\in \Sigma$, it follows that $(F^j(c)_i)_{i,j}$ is $\varnothing$-automatic.
\end{proof}

\section{Free Commutative  $(\iindex,\period)$-Monoids}\label{sec:free_monoids}

In this section, we will see how we can combine what we know from Proposition~\ref{prop:corollary} about groups with what we know from Proposition~\ref{prop:aperiodic} about aperiodic monoids.  We will start with an example in Section~\ref{sec:free_example}, which we will generalize in Section~\ref{sec:monogenic_monoids} to all (finite) monogenic monoids, before treating the case of all (finite) free commutative $(\iindex,\period)$-monoids in Section~\ref{sec:free_monoids_subsection}.
 
\subsection{Third Example}\label{sec:free_example}

Let $M =\left\langle a | a^6= a^4\right\rangle$, whose table is given in Figure~\ref{fig:table_1}.

\begin{figure}[htbp]
\centering
$\begin{array}{c|cccccc}
\cdot & 1 & a & a^2 & a^3 & a^4 & a^5 \\
\hline
1 & 1 & a &  a^2 &  a^3 & a^4 & a^5 \\
a & a & a^2 & a^3 & a^4 & a ^5 & a^4 \\
a^2 & a^2 & a^3 &  a^4 &  a^5 & a^4 & a^5 \\
a^3 & a^3 & a^4 & a^5 & a^4 & a ^5 & a^4 \\
a^4 & a^4 & a^5 &  a^4 &  a^5 & a^4 & a^5 \\
a^5 & a^5 & a^4 & a^5 & a^4 & a ^5 & a^4 \\
\end{array}$
\caption{Monoid table for $M$\label{fig:table_1}}
\end{figure}

Let $f_0=f_1=\id_M$.  $f_0$ and $f_1$ define a global transition function $F$ that is a cellular automaton and an endomorphism of $M^{\Z}$. 
When we run this cellular automaton on the initial configuration $\bar a$, we get Figure~\ref{fig:free_exemple_1}.

\begin{figure}[htbp]
\centering
\begin{tikzpicture}  [help lines/.style={draw=black},  every node/.style={help lines,rectangle,minimum size=6mm},  cellular automaton/.style={draw=none,row sep=0mm,column sep=0mm},  rst/.style={fill=white,help lines},  exc/.style={fill=blue!70,help lines},  ref/.style={fill=blue!30!white,help lines}]
    \matrix[cellular automaton] {
\node[rst] {$a$}; \\
\node[rst] {$a$};&\node[rst] {$a$};  \\
\node[rst] {$a$};&\node[rst] {$a^2$};  &\node[rst] {$a$};  \\
\node[rst] {$a$};&\node[rst] {$a^3$};  &\node[rst] {$a^3$}; & \node[rst] {$a$}; \\
\node[rst] {$a$};&\node[rst] {$a^4$};  &\node[rst] {$a^4$}; & \node[rst] {$a^4$};  &\node[rst] {$a$}; \\
\node[rst] {$a$};&\node[rst] {$a^5$};  &\node[rst] {$a^4$};  &\node[rst] {$a^4$}; & \node[rst] {$a^5$}; &\node[rst] {$a$}; \\
\node[rst] {$a$};&\node[rst] {$a^4$};  &\node[rst] {$a^5$};  &\node[rst] {$a^4$}; & \node[rst] {$a^5$}; &\node[rst] {$a^4$};  & \node[rst] {$a$}; \\
\node[rst] {$a$};&\node[rst] {$a^5$};  &\node[rst] {$a^5$};  &\node[rst] {$a^5$}; & \node[rst] {$a^5$}; &\node[rst] {$a^5$}; & \node[rst] {$a^5$};  & \node[rst] {$a$} ;  \\
\node[rst] {$a$};&\node[rst] {$a^4$};  &\node[rst] {$a^4$};  &\node[rst] {$a^4$}; & \node[rst] {$a^4$}; &\node[rst] {$a^4$}; & \node[rst] {$a^4$};  & \node[rst] {$a^4$};  & \node[rst] {$a$} ; \\
\node[rst] {$a$};&\node[rst] {$a^5$};  &\node[rst] {$a^4$};  &\node[rst] {$a^4$}; & \node[rst] {$a^4$}; &\node[rst] {$a^4$}; & \node[rst] {$a^4$};  & \node[rst] {$a^4$};  & \node[rst] {$a^5$};  & \node[rst] {$a$} ; \\
\node[rst] {$a$};&\node[rst] {$a^4$};  &\node[rst] {$a^5$};  &\node[rst] {$a^4$}; & \node[rst] {$a^4$}; &\node[rst] {$a^4$}; & \node[rst] {$a^4$};  & \node[rst] {$a^4$};  & \node[rst] {$a^5$};  & \node[rst] {$a^4$} ;  & \node[rst] {$a$} ; \\
    };
  \end{tikzpicture}
\caption{Ten iterations of $F$ on the initial configuration $\bar a$.  The top left cell has coordinates $(0,0)$; time flows downwards.  The neutral element $1$ is not depicted.\label{fig:free_exemple_1}}
\end{figure}

This is the image of Pascal's triangle under the morphism $\phi:\N\to M$ defined by $\phi(1) = a$.  A quick glance at it suggests that it can be understood by somehow separating $M$ into its ``aperiodic component" $\{1,a,a^2,a^3\}$ and its "periodic component" $\{a^4,a^5\}$, using the results from Sections \ref{sec:groups} and \ref{sec:aperiodic} to conclude, and this is more or less what we are going to do.


\subsubsection*{Aperiodic component}

If, to the presentation of $M$, we add the relation $a^5=a^4$, we get  $\left\langle a | a^6= a^4,a^5=a^4 \right\rangle =  \left\langle a | a^5= a^4\right\rangle\simeq (P_5,+,0)$, and with it comes a morphism $\alpha:M\to P_5$ defined by $\alpha(a)=1$.

The endomorphisms $g_0=g_1=\id_{P_5}$ define a global transition function $G$ that is a cellular automaton and an endomorphism of $P_5^{\Z}$, whose spacetime diagram, shown in Figure~ \ref{fig:free_exemple_2}, is the image of that of $F$ by  $\alpha$.  $G$ is the ``aperiodic component" of $F$.  Since $P_5$ is an aperiodic monoid, according to Proposition~\ref{prop:aperiodic}, this spacetime diagram is $\varnothing$-automatic.

\begin{figure}[htbp]
\centering
\begin{tikzpicture}  [help lines/.style={draw=black},  every node/.style={help lines,rectangle,minimum size=6mm},  cellular automaton/.style={draw=none,row sep=0mm,column sep=0mm},  rst/.style={fill=white,help lines},  exc/.style={fill=blue!70,help lines},  ref/.style={fill=blue!30!white,help lines}]
    \matrix[cellular automaton] {
\node[rst] {$1$}; \\
\node[rst] {$1$};&\node[rst] {$1$};  \\
\node[rst] {$1$};&\node[rst] {$2$};  &\node[rst] {$1$};  \\
\node[rst] {$1$};&\node[rst] {$3$};  &\node[rst] {$3$}; & \node[rst] {$1$}; \\
\node[rst] {$1$};&\node[rst] {$4$};  &\node[rst] {$4$}; & \node[rst] {$4$};  &\node[rst] {$1$}; \\
\node[rst] {$1$};&\node[rst] {$4$};  &\node[rst] {$4$};  &\node[rst] {$4$}; & \node[rst] {$4$}; &\node[rst] {$1$}; \\
\node[rst] {$1$};&\node[rst] {$4$};  &\node[rst] {$4$};  &\node[rst] {$4$}; & \node[rst] {$4$}; &\node[rst] {$4$};  & \node[rst] {$1$}; \\
\node[rst] {$1$};&\node[rst] {$4$};  &\node[rst] {$4$};  &\node[rst] {$4$}; & \node[rst] {$4$}; &\node[rst] {$4$}; & \node[rst] {$4$};  & \node[rst] {$1$} ;  \\
\node[rst] {$1$};&\node[rst] {$4$};  &\node[rst] {$4$};  &\node[rst] {$4$}; & \node[rst] {$4$}; &\node[rst] {$4$}; & \node[rst] {$4$};  & \node[rst] {$4$};  & \node[rst] {$1$} ; \\
\node[rst] {$1$};&\node[rst] {$4$};  &\node[rst] {$4$};  &\node[rst] {$4$}; & \node[rst] {$4$}; &\node[rst] {$4$}; & \node[rst] {$4$};  & \node[rst] {$4$};  & \node[rst] {$4$};  & \node[rst] {$1$} ; \\
\node[rst] {$1$};&\node[rst] {$4$};  &\node[rst] {$4$};  &\node[rst] {$4$}; & \node[rst] {$4$}; &\node[rst] {$4$}; & \node[rst] {$4$};  & \node[rst] {$4$};  & \node[rst] {$4$};  & \node[rst] {$4$} ;  & \node[rst] {$1$} ; \\
    };
  \end{tikzpicture}
\caption{Ten iterations of $G$ on the initial configuration $\bar 1$.  The top left cell has coordinates $(0,0)$; time flows downwards.  The neutral element $0$ is not depicted.\label{fig:free_exemple_2}}
\end{figure}

\subsubsection*{Periodic component}

Let us now modify the presentation of $M$ by adding the relation $a^4=1$: we get $\left\langle a | a^6= a^4,a^4=1 \right\rangle =  \left\langle a | a^2= 1\right\rangle\simeq \Z_2$; this yields a morphism $\beta:M\to (\Z_2,+,0)$ defined by $\beta(a)=1$.

The endomorphisms $h_0=h_1=\id_{\Z_2}$ define a global transition function $H$ that is a cellular automaton and an endomorphism of $\Z_2^{\Z}$, whose spacetime diagram, shown in Figure~\ref{fig:free_exemple_3},  is the image of that of $F$ by $\beta$.  $H$ is the ``periodic component" of $F$. Since $\Z_2$ is a group of order 2, according to Proposition~\ref{prop:corollary}, this spacetime diagram is 2-automatic.

\begin{figure}[htbp]
\centering
\begin{tikzpicture}  [help lines/.style={draw=black},  every node/.style={help lines,rectangle,minimum size=6mm},  cellular automaton/.style={draw=none,row sep=0mm,column sep=0mm},  rst/.style={fill=white,help lines},  exc/.style={fill=blue!70,help lines},  ref/.style={fill=blue!30!white,help lines}]
    \matrix[cellular automaton] {
\node[rst] {$1$}; \\
\node[rst] {$1$};&\node[rst] {$1$};  \\
\node[rst] {$1$};&\node[rst] {$0$};  &\node[rst] {$1$};  \\
\node[rst] {$1$};&\node[rst] {$1$};  &\node[rst] {$1$}; & \node[rst] {$1$}; \\
\node[rst] {$1$};&\node[rst] {$0$};  &\node[rst] {$0$}; & \node[rst] {$0$};  &\node[rst] {$1$}; \\
\node[rst] {$1$};&\node[rst] {$1$};  &\node[rst] {$0$};  &\node[rst] {$0$}; & \node[rst] {$1$}; &\node[rst] {$1$}; \\
\node[rst] {$1$};&\node[rst] {$0$};  &\node[rst] {$1$};  &\node[rst] {$0$}; & \node[rst] {$1$}; &\node[rst] {$0$};  & \node[rst] {$1$}; \\
\node[rst] {$1$};&\node[rst] {$1$};  &\node[rst] {$1$};  &\node[rst] {$1$}; & \node[rst] {$1$}; &\node[rst] {$1$}; & \node[rst] {$1$};  & \node[rst] {$1$} ;  \\
\node[rst] {$1$};&\node[rst] {$0$};  &\node[rst] {$0$};  &\node[rst] {$0$}; & \node[rst] {$0$}; &\node[rst] {$0$}; & \node[rst] {$0$};  & \node[rst] {$0$};  & \node[rst] {$1$} ; \\
\node[rst] {$1$};&\node[rst] {$1$};  &\node[rst] {$0$};  &\node[rst] {$0$}; & \node[rst] {$0$}; &\node[rst] {$0$}; & \node[rst] {$0$};  & \node[rst] {$0$};  & \node[rst] {$1$};  & \node[rst] {$1$} ; \\
\node[rst] {$1$};&\node[rst] {$0$};  &\node[rst] {$1$};  &\node[rst] {$0$}; & \node[rst] {$0$}; &\node[rst] {$0$}; & \node[rst] {$0$};  & \node[rst] {$0$};  & \node[rst] {$1$};  & \node[rst] {$0$} ;  & \node[rst] {$1$} ; \\
    };
  \end{tikzpicture}
\caption{Ten iterations of $H$ on the initial configuration $\bar 1$.  The top left cell has coordinates $(0,0)$; time flows downwards.  The neutral element $0$ is not depicted.\label{fig:free_exemple_3}}
\end{figure}

\subsubsection*{Conclusion}

A crucial property is that an element $x\in M$ can be recovered from $\alpha(x)$ and $\beta(x)$.  More formally, there exists a function $\gamma:P_5\times \Z_2\to M$ such that, for every $x\in M$, $\gamma(\alpha(x),\beta(x))= x$.   Namely, $\gamma(i,j)=\left\{ \begin{array}{ll}a^i & \text{if $i<4$} \\ a^{4+j} & \text{if i=4}  \end{array}\right.$.

It follows that the spacetime diagram produced by $F$ on the initial configuration $\bar a$ is 2-automatic.

\subsection{Monogenic Monoids}\label{sec:monogenic_monoids}

Monogenic monoids are monoids generated by a single element $a$.  Finite monogenic monoids are characterized by two integers: the threshold $\iindex\geq 0$ and the period $\period>0$ of $a$.
Let us denote $\freemonoid{\iindex}{\period} =  \left\langle a | a^{\iindex+\period} = a^{\iindex}\right \rangle$: $\iindex$  and $\period$ are referred to as the threshold and period of $\freemonoid{\iindex}{\period}$.
The following observation will be of use:

\begin{prop}\label{prop:monogenic_unique}
For every $k\in\N$, there exists a unique endomorphism $f$ of $\freemonoid{\iindex}{\period}$ such that $f(a)=a^k$.
\end{prop}

We define the aperiodic component of $\freemonoid{\iindex}{\period}$ by adding the relation $a^{\iindex+1}=a^\iindex$ to the presentation of $\freemonoid{\iindex}{\period}$: $\left\langle a | a^{\iindex+\period} = a^{\iindex}, a^{\iindex+1}=a^\iindex \right \rangle = \left\langle a | a^{\iindex+1} = a^{\iindex}\right \rangle = \freemonoid{\iindex}{1} \simeq P_{\iindex+1}$.  We thus get a morphism $\alpha: \freemonoid{\iindex}{\period}\to P_{\iindex+1}$.
Likewise, the periodic component of $\freemonoid{\iindex}{\period}$ is found by adding the relation $a^{\period}=1$ to the presentation of $\freemonoid{\iindex}{\period}$: $\left\langle a | a^{\iindex+\period} = a^{\iindex}, a^{\period}=1 \right \rangle = \left\langle a | a^{\period} = 1 \right \rangle = \freemonoid{0}{\period} \simeq \Z_p$,  which gives a morphism $\beta:\freemonoid{\iindex}{\period}\to \Z_\period$.

Let $F:\freemonoid{\iindex}{\period}^\Z\to \freemonoid{\iindex}{\period}^\Z$ be a cellular automaton that is also an endomorphism of $\freemonoid{\iindex}{\period}^\Z$. It is defined by a family $(f_i)_{i\in I}$ of endomorphisms of $\freemonoid{\iindex}{\period}$.   For each $i\in I$, according to Proposition~\ref{prop:monogenic_unique}, there exist unique endomorphisms $g_i$ and $h_i$ of, respectively, $P_{\iindex+1}$ and $\Z_{\period}$, such that $g_i(1)=\alpha(f_i(a))$ and $h_i(1)=\beta(f_i(a))$.  Let $x\in \freemonoid{\iindex}{\period}$, and let $k\in\N$ be such that $x=a^k$. We have $g_i(\alpha(a^k))=g_i(k.1) =k.g_i(1) = k.\alpha(f_i(a)) = \alpha(f_i(a)^k) = \alpha(f_i(a^k))$; therefore $\alpha\circ f_i = g_i\circ\alpha$.
Likewise, $\beta\circ f_i=h_i\circ\beta$.

These families of endomorphisms thus define cellular automata $G$ and $H$ that are endomorphisms of, respectively, $P_{\iindex+1}^\Z$ and $\Z_{\period}^\Z$, such that
$\alpha\circ F =  G\circ\alpha$ and $\beta\circ F = H\circ\beta$.  
Let $c\in \freemonoid{\iindex}{\period}^\Z$ be a finite configuration.  
For any $n\in\N$, $\alpha(F^n(c)) = G^n(\alpha(c))$ and $\beta(F^n(c))=H^n(\beta(c))$ (where, with some abuse of notation, we also denote by $\alpha$ and $\beta$ their simultaneous application to every cell of a configuration).

We thus split our spacetime diagram into an aperiodic and a periodic component.  According to Proposition~\ref{prop:aperiodic}, the spacetime diagram produced by $G$ with the initial configuration $\alpha(c)$ is $\varnothing$-automatic.  According to Proposition~\ref{prop:corollary}, the spacetime diagram produced by $H$ with the initial configuration $\beta(c)$ is $\pi(\Z_\period)$-automatic.

Let $\gamma:P_{\iindex+1}\times\Z_\period \to \freemonoid{\iindex}{\period}$ be the function (it is not a morphism!) defined by $\gamma(k,j) = \left\{ \begin{array}{ll}a^k & \text{if $k<\iindex$} \\ a^{\iindex\period+j} & \text{if $ k=\iindex$}  \end{array}\right.$.  Then, for all $x\in \freemonoid{\iindex}{\period}$, $\gamma(\alpha(x),\beta(x))=x$.  Therefore the spacetime diagram produced by $F$ with the initial configuration $c$ is $\pi(\Z_\period)=\pi(\freemonoid{\iindex}{\period})$-automatic.

\subsection{Free Commutative $(\iindex,\period)$-Monoids} \label{sec:free_monoids_subsection}

Let $\iindex$ be a nonnegative integer, $\period$ and $r$ positive integers.  The free commutative $(\iindex,\period)$-monoid of rank $r$ is  $\freemonoid{\iindex}{\period}^r = (\freemonoid{\iindex}{\period})^r$.

Let $\alpha:\freemonoid{\iindex}{\period}\to P_{\iindex+1}$, $\beta:\freemonoid{\iindex}{\period}\to \Z_\period$ be the morphisms defined in Section~\ref{sec:monogenic_monoids}.
We define the morphisms $\alpha_r: \freemonoid{\iindex}{\period}^r\to P_{\iindex+1}^r$ and $\beta_r: \freemonoid{\iindex}{\period}^r\to \Z_{\period}^r$ by:

\begin{equation}
\begin{split}
\alpha_r(x_1,\ldots,x_r)&= (\alpha (x_1),\ldots,\alpha(x_r)) \\ 
 \beta_r(x_1,\ldots,x_r)&=(\beta(x_1),\ldots, \beta(x_r))
\end{split}
\end{equation}

Likewise, we define  $\gamma_r:P_{\iindex+1}^r\times\Z_\period^r \to \freemonoid{\iindex}{\period}^r$ by

\begin{equation}
\gamma_r((k_1,\ldots,k_r),(j_1,\ldots,j_r)) =(\gamma(k_1,j_1),\ldots,\gamma(k_r,j_r)).
\end{equation}

Then, for all $x\in \freemonoid{\iindex}{\period}^r$, $\gamma_r(\alpha_r(x),\beta_r(x))=x$.  For $n\in\intint{1;r}$, let us denote by $a_n$ a generator of the $n$-th copy of $\freemonoid{\iindex}{\period}$ in $\freemonoid{\iindex}{\period}^r$, i.e., $a_n = \gamma_r((k_1,\ldots,k_r),(j_1,\ldots,j_r))$, where $k_i=j_i = \left\{\begin{array}{ll}
1 &  \text{if $i=n$} \\
0 & \text{if $i\neq n$} \\
\end{array}\right.$.   Like in the monogenic case, endomorphisms of $\freemonoid{\iindex}{\period}^r$ are fairly easy to describe:

\begin{prop}\label{prop:polygenic_unique}
For every $x_1,\ldots,x_r \in \freemonoid{\iindex}{\period}^r$, there exists a unique endomorphism $f$ of $\freemonoid{\iindex}{\period}^r$ such that, for every $j\in\intint{1;r}$,  $f(a_j)=x_j$.
\end{prop}

It follows that the same technique applied in the monogenic case also applies in the more general case of free commutative $(\iindex,\period)$-monoids.
Let $F:(\freemonoid{\iindex}{\period}^r)^\Z\to (\freemonoid{\iindex}{\period}^r)^\Z$ be a cellular automaton that is also an endomorphism of $(\freemonoid{\iindex}{\period}^r)^\Z$. It is defined by a family $(f_i)_{i\in I}$ of endomorphisms of $\freemonoid{\iindex}{\period}^r$. 

For each $i\in I$, according to Proposition~\ref{prop:polygenic_unique}, there exist unique endomorphisms $g_i$ and $h_i$ of, respectively, $P_{\iindex+1}^r$ and $\Z_{\period}^r$, such that for every $j\in \intint{1;r}$,  $g_i(\alpha_r(a_j))=\alpha_r(f_i(a_j))$ and $h_i(\beta_r(a_j))=\beta_r(f_i(a_j))$.   We then have $\alpha_r\circ f_i = g_i\circ\alpha_r$ and $\beta_r\circ f_i=h_i\circ\beta_r$.

These families of endomorphisms define cellular automata $G$ and $H$ that are endomorphisms of, respectively, $(P_{\iindex+1}^r)^\Z$ and $(\Z_{\period}^r)^\Z$, such that
$\alpha_r\circ F =  G\circ\alpha_r$ and $\beta_r\circ F = H\circ\beta_r$.  
Let $c\in(\freemonoid{\iindex}{\period}^r)^Z$ be a finite configuration.  
For any $n\in\N$, $\alpha_r(F^n(c)) = G^n(\alpha_r(c))$ and $\beta_r(F^n(c))=H^n(\beta_r(c))$.

According to Proposition~\ref{prop:aperiodic}, the spacetime diagram produced by $G$ with the initial configuration $\alpha_r(c)$ is $\varnothing$-automatic.
According to Proposition~\ref{prop:corollary}, the spacetime diagram produced by $H$ with the initial configuration $\beta_r(c)$ is $\pi(\Z_\period^r)$-automatic.  Notice that $\pi(\Z_\period^r) = \pi(\freemonoid{\iindex}{\period}^r)$ is the set of primes dividing $\period$.

Since the spacetime diagram produced by $F$ on the initial configuration $c$ is the image of those two diagrams by the function $\gamma_r$, it is also $\pi(\freemonoid{\iindex}{\period}^r)$-automatic.

\section{General Case} \label{sec:general_case}
In Section~\ref{sec:free_monoids}, we proved that Theorem~\ref{thm:main} is true when $\Sigma$ is a free commutative $(\iindex,\period)$-monoid.   The aim of this section is to justify that  free commutative $(\iindex,\period)$-monoids essentially encompass all the complexity that can be encountered when $\Sigma$ is an arbitrary commutative monoid --- much like in \cite{Gutschow2010fractal} the general case of abelian groups was reduced to the study of $\Z_{\period}^r$.   More precisely, we will show that spacetime diagrams produced by cellular automata on commutative monoids are projections of spacetime diagrams produced by cellular automata on free commutative $(\iindex,\period)$-monoids.

\begin{prop}\label{prop:covering_free_monoid}
Let $M$ be a finite commutative monoid. Let $\iindex$  be the maximum threshold of the elements of $M$, and  $\period$ the least common multiple of their periods. There exists an integer $r$ and a surjective morphism  $\phi:\freemonoid{\iindex}{\period}^r \to M$ such that for any endomorphism $f$ of $M$,  there exists an endomorphism $\tilde f$ of $\freemonoid{\iindex}{\period}^r$, such that the following diagram commutes:

\begin{displaymath}
  \begin{tikzcd}
     \freemonoid{\iindex}{\period}^r \arrow{r}{\tilde f} \arrow[swap, two heads]{d}{\phi} &   \freemonoid{\iindex}{\period}^r  \arrow[two heads]{d}{\phi} \\
    M \arrow{r} {f} & M
  \end{tikzcd}
\end{displaymath}

\end{prop}

\begin{proof}
Let $\left\langle X | R\right\rangle$ be a presentation of $M$: $X$ is a finite set of generators, and $R$ is a set of relations on $ X^*$.   

Let $x,y\in M$, and $i,j,p,q$ positive integers such that $x^{i+p}=x^i$ and $y^{j+q}=y^j$.  Let $k=\max(i,j)$ and $s=\lcm(p, q)$.   Then
$x^{k+s} = x^{i+(k-i) + p \times \frac{s}{p}} = x^{k-i} x^{i + \frac{s}{p} \times  p} =  x^{k-i} x^{i} =x^k$.   Therefore, for every $x\in M$, $x^{\iindex+\period}=x^\iindex$, where $\iindex$ is the maximum threshold of the elements of $M$ and $\period$ the least common multiple of their periods.

Let $E = \{xy=yx | x,y\in X\}\cup \{x^{\iindex+\period}=x^{\iindex} |x \in X\}$.   
Since $M$ is a commutative $(\iindex,\period)$-monoid, $\left\langle X | R\cup E\right\rangle=\left\langle X | R\right\rangle=M$. 
Let $r = |X|$ and $\phi$ be the projection morphism  $\phi: \freemonoid{\iindex}{\period}^r \simeq \left\langle X | E\right\rangle  \to    \left\langle X | R\cup E \right\rangle = M$.

Let $f$ be an endomorphism of $M$.  According to Proposition~\ref{prop:polygenic_unique},  there exists an endomorphism $\tilde f$ of $ \left\langle X | E\right\rangle $ such that for every $x\in X$, $ \phi(\tilde f(x))=  f(\phi(x))$: just define $\tilde f(x)$ to be any element of $\phi^{-1}(f(\phi(x)))$.  Then, by construction, $\phi \circ \tilde f = f \circ \phi$.
\end{proof}

We can now complete the proof of Theorem~\ref{thm:main}.  Let $\Sigma$ be a finite commutative monoid, and $c\in \Sigma^\Z$ a finite configuration. Let $F:\Sigma^\Z\to\Sigma^\Z$ be a cellular automaton that is also an endomorphism of $\Sigma^\Z$, and let $(f_i)_{i\in I}$ be the corresponding family of local endomorphisms.

According to Proposition~\ref{prop:covering_free_monoid}, there are integers $\iindex$, $\period$, $r$, a surjective morphism $\phi:\freemonoid{\iindex}{\period}^r \to \Sigma$ and a family of endomorphisms $(\tilde f_i)_{i\in I}$ of $\freemonoid{\iindex}{\period}^r$ such that, for every $i\in I$, $\phi\circ \tilde f_i = f_i \circ \phi$.  These in turn define a global transition function $\tilde F:(\freemonoid{\iindex}{\period}^r)^\Z \to (\freemonoid{\iindex}{\period}^r)^\Z$.

Let $\tilde c$ be a finite configuration on the alphabet $\freemonoid{\iindex}{\period}^r$ such that $\phi(\tilde c)= c$.
For any $(i,j)\in \Z\times\N$,

\begin{align*}
\phi (\tilde F^j(\tilde c)_i) &= \phi\left(\prod\limits_{\substack{x_0\in \supp(\tilde c) \\ x_1,\ldots,x_j \in I  \\ x_0+\cdots+x_j = i}}  \tilde f_{x_j}\tilde f_{x_{j-1}}\cdots \tilde f_{x_1}(\tilde c_{x_0})\right)\\
&=  \prod\limits_{\substack{x_0\in \supp(\tilde c) \\ x_1,\ldots,x_j \in I  \\ x_0+\cdots+x_j = i}}  \phi \tilde f_{x_j} \tilde f_{x_{j-1}}\cdots \tilde f_{x_1}(\tilde c_{x_0})\\
& =  \prod\limits_{\substack{x_0\in \supp(\tilde c) \\ x_1,\ldots,x_j \in I  \\ x_0+\cdots+x_j = i}}     f_{x_j}\phi \tilde f_{x_{j-1}}\cdots \tilde f_{x_1}(\tilde c_{x_0})\\
& =  \prod\limits_{\substack{x_0\in \supp(\tilde c) \\ x_1,\ldots,x_j \in I  \\ x_0+\cdots+x_j = i}}     f_{x_j} f_{x_{j-1}}\cdots  f_{x_1}\phi(\tilde c_{x_0})\\
& =  \prod\limits_{\substack{x_0\in \supp(c) \\ x_1,\ldots,x_j \in I  \\ x_0+\cdots+x_j = i}}     f_{x_j} f_{x_{j-1}}\cdots  f_{x_1}(c_{x_0})\\
&  = F^j(c)_i
\end{align*}

The spacetime diagram of $F$ is thus the image of that of $\tilde F$ by $\phi$.  But we saw in Section~\ref{sec:free_monoids} that the spacetime diagram of $\tilde F$, on the initial configuration $\tilde c$, is $A$-automatic, where $A$ is the set of primes dividing $\period$. Since $\period$ is the least common multiple of the periods of the elements of $\Sigma$, then $A=\pi(\Sigma)$; therefore the spacetime diagram of $F$ on the initial configuration $c$ is $\pi(\Sigma)$-automatic.  This concludes the proof of Theorem~\ref{thm:main}.

\section{Automatic initial configuration} \label{sec:automatic_initial}

This section is devoted to the proof of Theorem~\ref{thm:automatic_initial_configuration}.  It was proven in \cite{rowland2020automaticity} in the case $\Sigma=\Z_p$, using Salon's theorem from \cite{salon1986suites}.  Our algebraic structures are apparently too weak to invoke such a powerful high-level theorem.   The idea sustaining our proof is very low-level: it works directly on the definition of $k$-automaticity.  We will show how to combine automata describing the spacetime diagrams of a cellular automaton on finite configurations with an automaton describing an initial configuration to derive an automaton describing the spacetime diagram starting on this automatic configuration.

\subsection{Fourth Example}

Arguably, the simplest nontrivial illustration of Theorem~\ref{thm:automatic_initial_configuration} has already been given in \cite{rowland2020automaticity}: It is Pascal's triangle (the Ledrappier cellular automaton) modulo 2, with an initial configuration that is the Prouhet-Thue-Morse sequence.

In order to illustrate our new result, one may think of the automaton $\Theta$ studied in \cite{Gutschow2010fractal}, whose alphabet is $\Z_2^2$, but the complexity of the description of the 2-automaticity of its spacetime diagram on an initial configuration even as simple as the Prouhet-Thue-Morse sequence is enormous.  In fact, it is even difficult to estimate the number of states of the automaton tasked to describe it.  If you follow to a T the procedure described in the following proof of Theorem~\ref{thm:automatic_initial_configuration}, you get perhaps $2^{\sim 200}$ states, although this number might in practice be lowered to a few hundred.  Actually, the automaton $\Gamma$ defined in \cite{NT2011Selfsimilarity} is probably a better candidate for a new example\ldots but for now, let us stick to a simpler one.  Let $U:\Z^2\to \Z_2$ be defined by:

\begin{itemize}
\item if $x<0$ or $y<0$, then $U(x,y)=0$;
\item $U(0,0)=0$;
\item for every $x\in \N$, $U(2x,0)=U(x,0)$ and $U(2x+1,0)=U(x,0)+1$
\item for every $x,y\in\N$, $U(x,y+1)=U(x,y)+U(x-1,y)$.
\end{itemize}

\begin{figure}[htbp]
\centering
\includegraphics[width=0.9\textwidth]{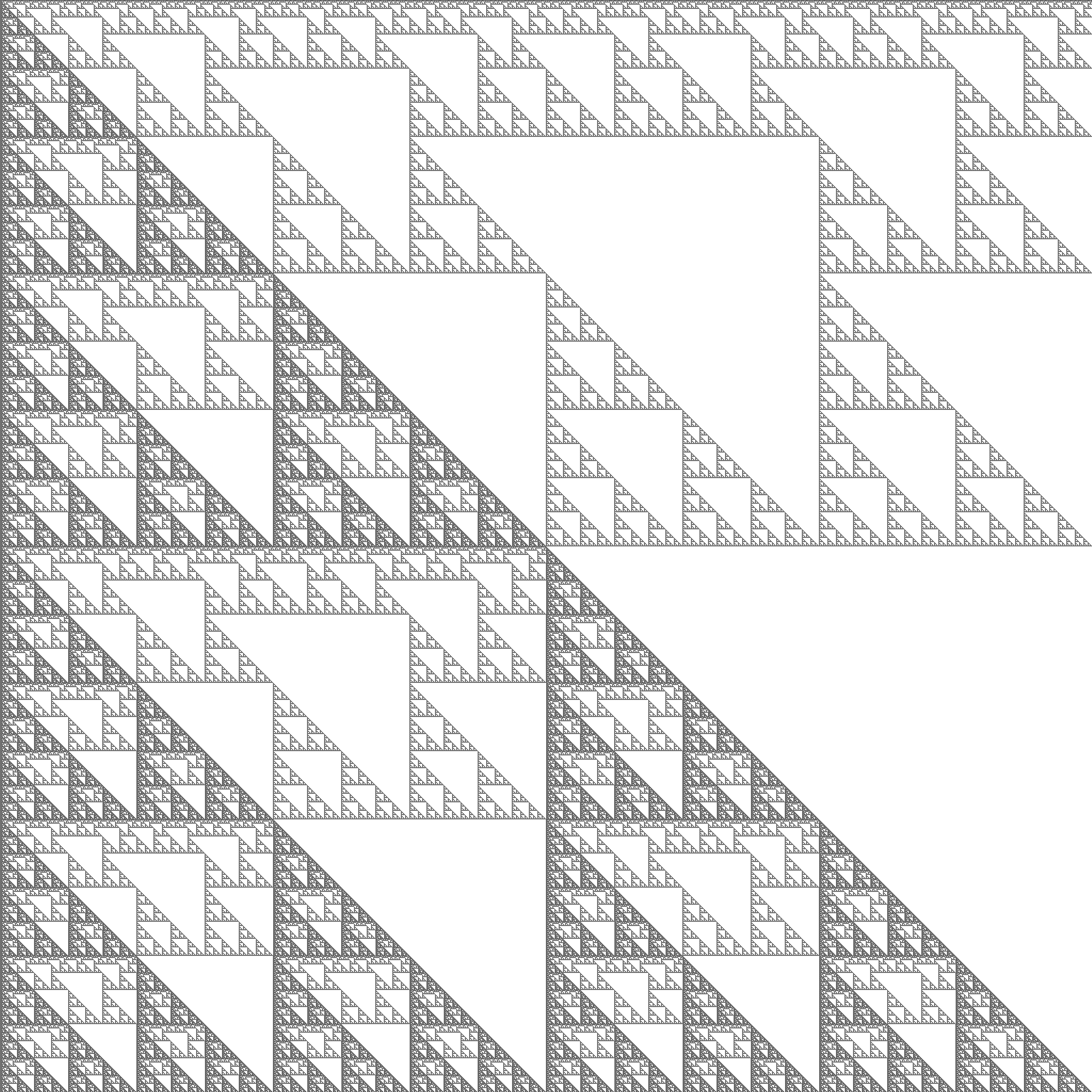}
\caption{The spacetime diagram on $\intint{0,2^{12}-1}^2$ of Pascal's triangle modulo 2 with the Prouhet-Thue-Morse sequence as initial configuration.\label{fig:PTM_Ledrappier}}
\end{figure}

$(U(x,0))_{x\geq 0}$ is the Prouhet-Thue-Morse sequence, and for $y\geq 0$, $(U(x,y+1))_{x\geq 0}$ is the image of $(U(x,y))_{x\geq 0}$ under the Ledrappier cellular automaton on $\Z_2$, i.e., Pascal's triangle rule modulo 2.   The spacetime diagram of $U$ up to a large power of 2 is represented in Figure~\ref{fig:PTM_Ledrappier}.

As in \cite{Gutschow2010fractal} and \cite{NT2011Selfsimilarity}, we will use the language of matrix substitution systems to describe what we are doing.
When we write 
\begin{displaymath}
\alpha \rightarrow \begin{array}{|c|c|}\hline \beta & \gamma \\ \hline \delta & \alpha  \\ \hline  \end{array}  \qquad  \alpha\mapsto 4
\end{displaymath}

 it means, in Definition~\ref{def:k-automatic}:
\begin{itemize}
\item $\alpha$, $\beta$, $\gamma$ and $\delta$ are elements of $E$;
\item if $e(x,y)=\alpha$, then $e(2x,2y)=\beta$, $e(2x+1,2y)=\gamma$, $e(2x,2y+1)=\delta$ and $e(2x+1,2y+1)=\alpha$;
\item if $e(x,y)=\alpha$ then $U(x,y)=4$.
\end{itemize}

In this fashion, here is a substitution system for the Prouhet-Thue-Morse sequence:

\begin{gather*}
\alpha \rightarrow \begin{array}{|c|c|}\hline \alpha & \beta \\ \hline \end{array} \qquad \beta \rightarrow \begin{array}{|c|c|}\hline \beta & \alpha \\  \hline \end{array} 
\quad  \qquad  \bot \rightarrow \begin{array}{|c|c|}\hline \bot & \bot \\  \hline \end{array} \\
\alpha \mapsto 0 \qquad \beta \mapsto 1   \qquad \bot \mapsto 0
\end{gather*}

It effectively defines this sequence if we add the information that the initial state for this substitution system is  $\begin{array}{|c|c|}\hline \bot&\alpha \\ \hline\end{array}$, meaning, in terms of Definition~\ref{def:k-automatic}, that $e(-1)=\bot$ and $e(0)=\alpha$.

Likewise, the spacetime diagram of Pascal's triangle modulo 2 can be thus described:

\begin{gather*}
A \rightarrow \begin{array}{|c|c|}\hline A & \times \\ \hline A & A \\ \hline  \end{array} \qquad \times \rightarrow \begin{array}{|c|c|}\hline \times & \times \\ \hline \times & \times \\ \hline  \end{array} \\
A \mapsto 1 \qquad   \qquad \times \mapsto 0
\end{gather*}

with the initial state being $\begin{array}{|c|c|}\hline \times & \times \\ \hline \times & A \\ \hline  \end{array}$, i.e., $A$ at coordinates $(0,0)$, $\times$ at $(-1,-1)$, $(-1,0)$ and $(0,-1)$.

Now, in order to find a matrix substitution system for $U$, we are going to combine both of these substitution systems.  Start from cell $(0,0)$.  There, the substitution system for the Prouhet-Thue-Morse sequence is in the state $\alpha$, while the one from Pascal's triangle is in the state $A$.  Let us signify both with a state denoted $A_\alpha$.  Since $\alpha$ evaluates to 0, we also have $A_\alpha\mapsto 0$.  What should we substitute it with?  $A$ is going to be substituted with $\begin{array}{|c|c|}\hline A & \times \\ \hline A & A \\ \hline  \end{array}$, so we might think the rule will be 
$A_\alpha\rightarrow\begin{array}{|c|c|}\hline A_\alpha & \times_\alpha \\ \hline A_\alpha & A_\alpha \\ \hline  \end{array}$.  But simultaneously, $\alpha$ is going to be substituted with  $\begin{array}{|c|c|}\hline \alpha & \beta \\ \hline \end{array}$, and we have to take that into account.   So, what we get is actually:

\def\lar{0.35 cm}
\begin{displaymath}
A_\alpha \rightarrow \begin{array}{|c|c|}\hline A_\alpha & \times_\alpha \\ \hline A_\alpha & A_\alpha  \\ \hline  \end{array}  + \begin{array}{|p{\lar}|c|c} \cline{1-2} &A_\beta & \times_\beta \\ \cline{1-2} & A_\beta &A_\beta \\ \cline{1-2}  \end{array} 	 = \begin{array}{|c|c|c} \cline{1-2} A_\alpha & A_\beta \times_\alpha & \times_\beta \\ \cline{1-2} A_\alpha & A_\alpha A_\beta & A_\beta \\ \cline{1-2}  \end{array} 	
\end{displaymath}

Since the alphabet of the cellular automaton is the group $\Z_2$, the sum is to be understood modulo 2, so we are defining a substitution system whose states are formal sums modulo 2 of elements of $\{\alpha,\beta,\bot\}\times \{A,\times\}$;  we denote this formal sum by juxtaposition: for instance, $A_\beta\times_\alpha$ is the formal sum modulo 2 of $A_\beta$ and $\times_\alpha$.  Likewise, we get:

\begin{displaymath}
A_\beta \rightarrow \begin{array}{|c|c|c} \cline{1-2} A_\beta & A_\alpha \times_\beta & \times_\alpha \\ \cline{1-2} A_\beta & A_\alpha A_\beta & A_\alpha \\ \cline{1-2}  \end{array} 	 \quad A_\bot \rightarrow \begin{array}{|c|p{\lar}|c} \cline{1-2} A_\bot &  & \times_\bot \\ \cline{1-2} A_\bot &  & A_\bot \\ \cline{1-2}  \end{array} 	 \quad 
\end{displaymath}

There is a small problem: it's not a substitution system yet, since it ``overflows" to the right.  There is a simple fix: grouping.  We add information to the state: we indicate in square brackets the state of the neighbor cell to the left.  The state of cell $(0,0)$ is thus $[A_\bot]A_\alpha$, and its substitution rule is:

\begin{displaymath}
[A_\bot]A_\alpha \rightarrow \begin{array}{|c|c|}\hline [] A_\alpha\times_\bot & [A_\alpha\times_\bot] A_\beta \times_\alpha \\ \hline [] A_\alpha A_\bot &  [A_\alpha A_\bot] A_\alpha A_\beta \\ \hline  \end{array} 
\end{displaymath}

We now have a substitution system --- a real one, this time --- whose set of states is $\Z_2^{(\{\alpha,\beta,\bot\}\times \{A,\times\})^2}$.  Fortunately, it can be greatly simplified: since both $\bot$ and $\times$ will always ultimately be evaluated to 0, they can be kept out.  The previous rule becomes:

\begin{displaymath}
[]A_\alpha \rightarrow \begin{array}{|c|c|}\hline [] A_\alpha & [A_\alpha] A_\beta  \\ \hline [] A_\alpha &  [A_\alpha ] A_\alpha A_\beta \\ \hline  \end{array} 
\end{displaymath}

If we note $a=A_\alpha$ and $b=A_\beta$, we get:

\begin{displaymath}
[]a \rightarrow \begin{array}{|c|c|}\hline [] a & [a] b \\ \hline [] a &  [a ] ab \\ \hline  \end{array} 
\end{displaymath}

So finally, we get our substitution system in its final form.  The set of states is the $\Z_2$-vector space generated by the basis $\{[a],[b],[]a,[]b\}$, and the substitutions are linear maps. They act on the basis states in this way:

\begin{displaymath}
[a]\rightarrow \begin{array}{|c|c|} \hline [b] & [] \\ \hline [ab]b & [b] \\ \hline \end{array} \quad [b]\rightarrow \begin{array}{|c|c|} \hline [a] & [] \\ \hline [ab]a & [a] \\ \hline \end{array} \quad []a \rightarrow \begin{array}{|c|c|}\hline [] a & [a] b \\ \hline [] a &  [a ] ab \\ \hline  \end{array} \quad  []b\rightarrow \begin{array}{|c|c|} \hline []b & [b]a \\ \hline []b & [b]ab \\ \hline \end{array} 
\end{displaymath}

\begin{figure}[htbp]
\begin{align*}
[]\rightarrow \begin{array}{|c|c|} \hline []  & [] \\ \hline [] & [] \\ \hline \end{array}  \qquad 
&  [b]\rightarrow \begin{array}{|c|c|} \hline [a] & [] \\ \hline [ab]a & [a] \\ \hline \end{array} \\
 []a\rightarrow \begin{array}{|c|c|} \hline []a & [a]b \\ \hline []a & [a]ab \\ \hline \end{array}  \qquad 
& [b]a\rightarrow \begin{array}{|c|c|} \hline [a]a & [a]b \\ \hline [ab] & []ab \\ \hline \end{array}  \\
 [a]\rightarrow \begin{array}{|c|c|} \hline [b] & [] \\ \hline [ab]b & [b] \\ \hline \end{array}  \qquad 
& [ab]\rightarrow \begin{array}{|c|c|} \hline [ab] & [] \\ \hline []ab & [ab] \\ \hline \end{array} \\
 [a]a\rightarrow \begin{array}{|c|c|} \hline [b]a & [a]b \\ \hline [ab]ab & [ab]ab \\ \hline \end{array}  \qquad 
& [ab]a\rightarrow \begin{array}{|c|c|} \hline [ab]a & [a]b \\ \hline []b & [b]ab \\ \hline \end{array} \\
\end{align*}
\caption{Substitution rules for the states evaluating to 0}
\label{fig:substitution_system_1}
\end{figure}

\begin{figure}[htbp]
\begin{align*}
[]b\rightarrow \begin{array}{|c|c|} \hline []b & [b]a \\ \hline []b & [b]ab \\ \hline \end{array} \qquad 
& [b]b\rightarrow \begin{array}{|c|c|} \hline [a]b & [b]a \\ \hline [ab]ab & [ab]ab \\ \hline \end{array} \\
[]ab\rightarrow \begin{array}{|c|c|} \hline []ab & [ab]ab \\ \hline []ab & [ab] \\ \hline \end{array} \qquad 
& [b]ab\rightarrow \begin{array}{|c|c|} \hline [a]ab & [ab]ab \\ \hline [ab]b & [b] \\ \hline \end{array} \\
[a]b\rightarrow \begin{array}{|c|c|} \hline [b]b & [b]a \\ \hline [ab] & []ab \\ \hline \end{array} \qquad 
& [ab]b\rightarrow \begin{array}{|c|c|} \hline [ab]b & [b]a \\ \hline []a & [a]ab \\ \hline \end{array} \\ 
[a]ab\rightarrow \begin{array}{|c|c|} \hline [b]ab & [ab]ab \\ \hline [ab]a & [a] \\ \hline \end{array} \qquad
& [ab]ab\rightarrow \begin{array}{|c|c|} \hline [ab]ab & [ab]ab \\ \hline [] & [] \\ \hline \end{array} \\
\end{align*}
\caption{Substitution rules for the states evaluating to 1}
\label{fig:substitution_system_2}
\end{figure}

From this we can deduce the whole substitution system (Figures~\ref{fig:substitution_system_1} and \ref{fig:substitution_system_2}).  We know the state at coordinates $(0,0)$ is $[]a$; it just remains to specify what all these states are going to evaluate to.  We have $[a],[b],[]a \mapsto 0$ and $[]b\mapsto 1$ and the evaluation function is linear; so, in Figure~\ref{fig:substitution_system_1}, all states evaluate to 0, while they all evaluate to 1 in Figure~\ref{fig:substitution_system_2}.  Let us say we want to compute $U(10,13)$.  In terms of Definition~\ref{def:k-automatic}, we have:
\begin{itemize}
\item  $e(0,0)=[]a$; 
\item  since $[]a\rightarrow \begin{array}{|c|c|} \hline []a & [a]b \\ \hline []a & [a]ab \\ \hline \end{array}$, $e(1,1)=[a]ab$;
\item since $[a]ab\rightarrow \begin{array}{|c|c|} \hline [b]ab & [ab]ab \\ \hline [ab]a & [a] \\ \hline \end{array}$, $e(2,3)=[ab]a$;
\item since $[ab]a\rightarrow \begin{array}{|c|c|} \hline [ab]a & [a]b \\ \hline []b & [b]ab \\ \hline \end{array}$, $e(5,6)=[a]b$;
\item since $[a]b\rightarrow \begin{array}{|c|c|} \hline [b]b & [b]a \\ \hline [ab] & []ab \\ \hline \end{array}$, $e(10,13)=[ab]$.
\end{itemize}

Since $[ab]\mapsto 0$, we conclude that $U(10,13)=0$.

\begin{figure}[htbp]
    \centering
    \begin{tabular}{cccc} 
        \subcaptionbox{$[]$\label{fig:1}}{\includegraphics[width=0.21\textwidth]{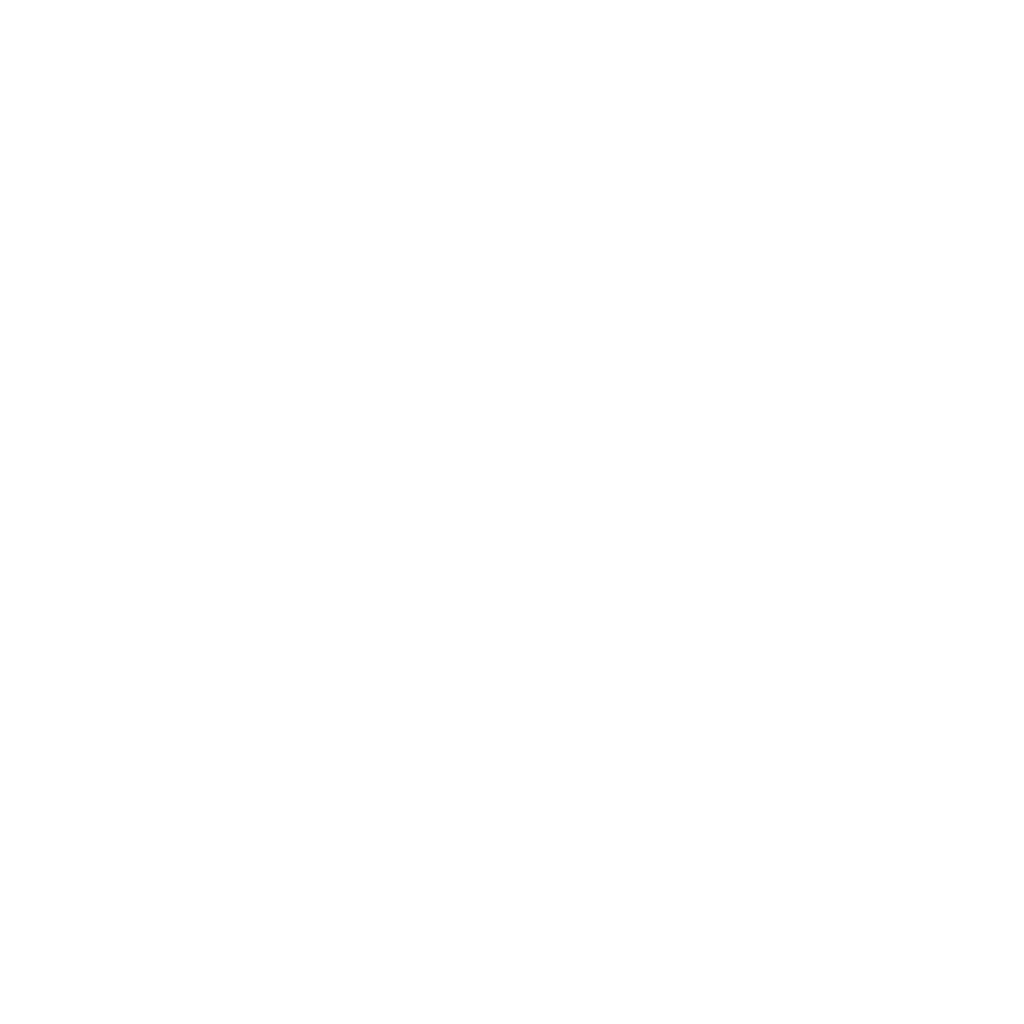}} &
        \subcaptionbox{$[]a$\label{fig:2}}{\includegraphics[width=0.21\textwidth]{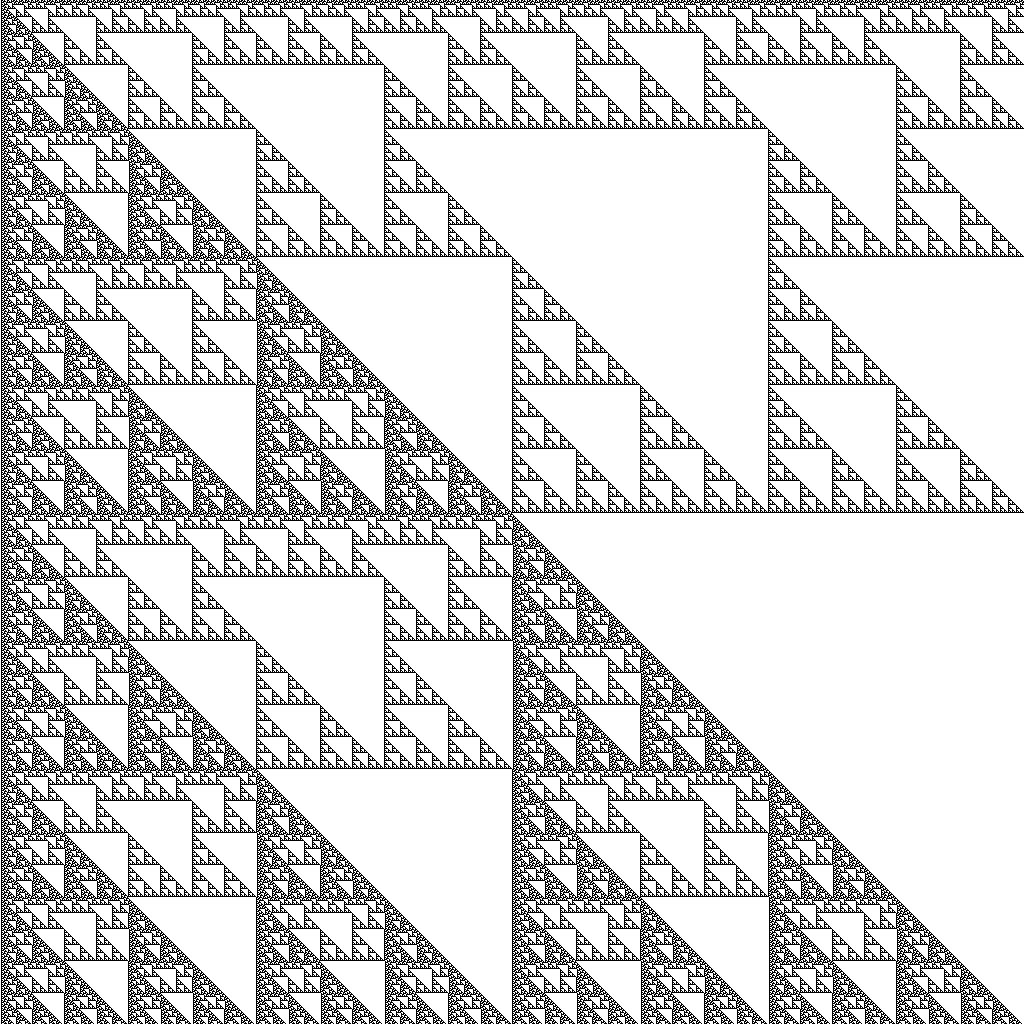}} &
        \subcaptionbox{$[]b$\label{fig:3}}{\includegraphics[width=0.21\textwidth]{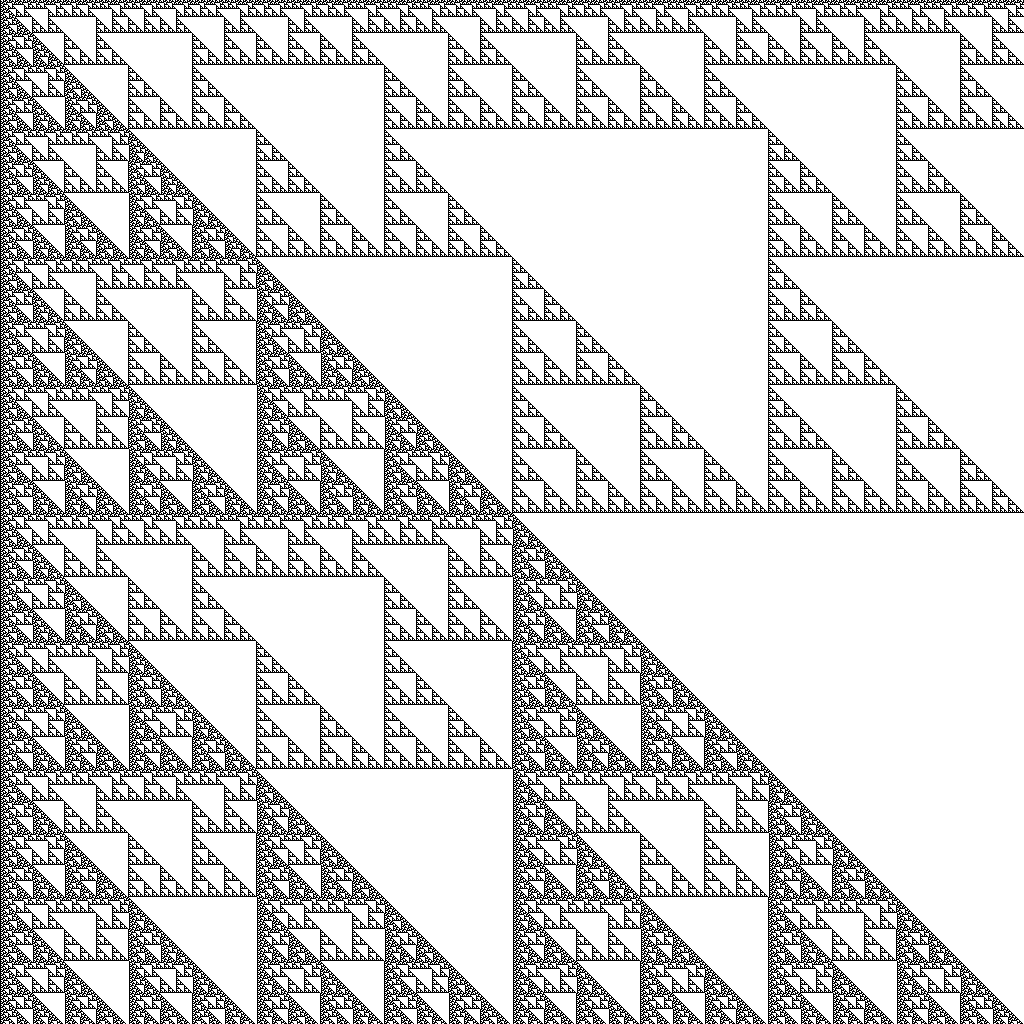}} &
        \subcaptionbox{$[]ab$\label{fig:4}}{\includegraphics[width=0.21\textwidth]{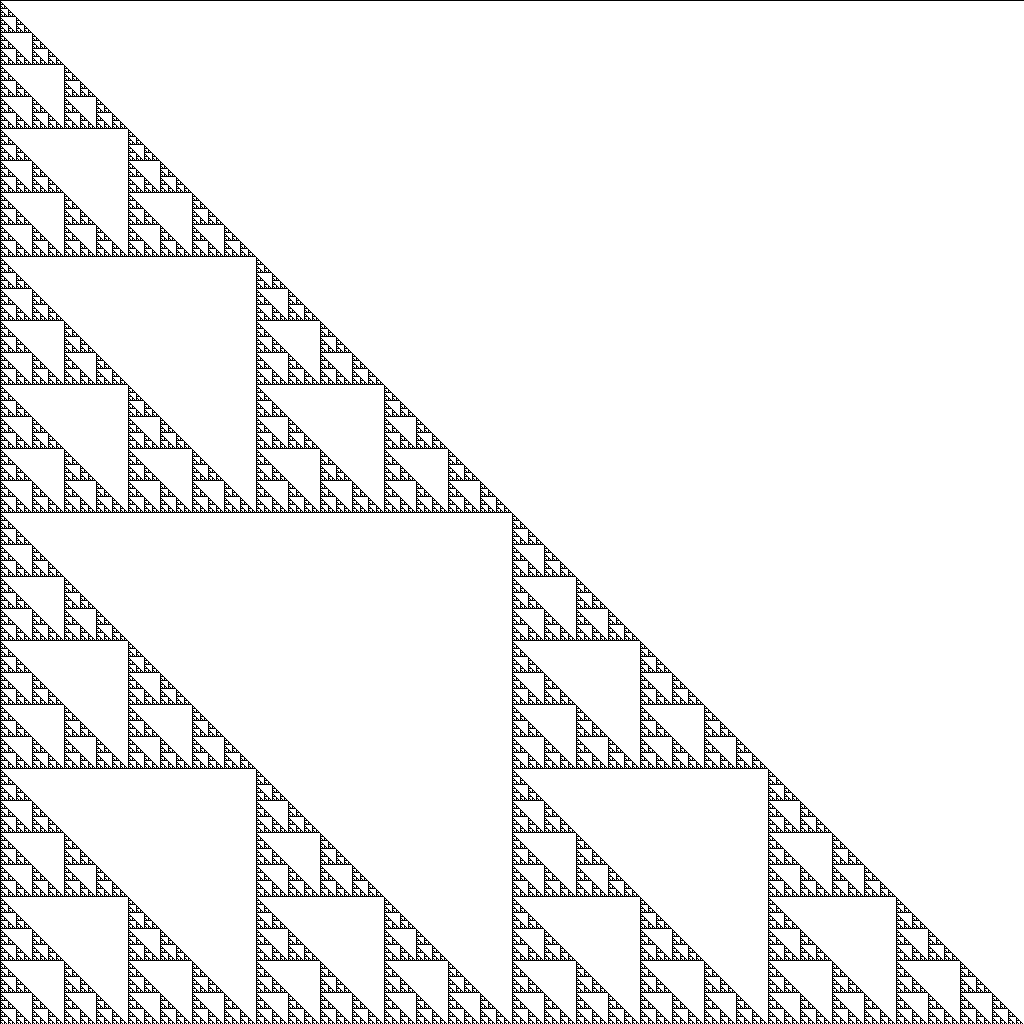}} \\
        
        \subcaptionbox{$[a]$\label{fig:5}}{\includegraphics[width=0.21\textwidth]{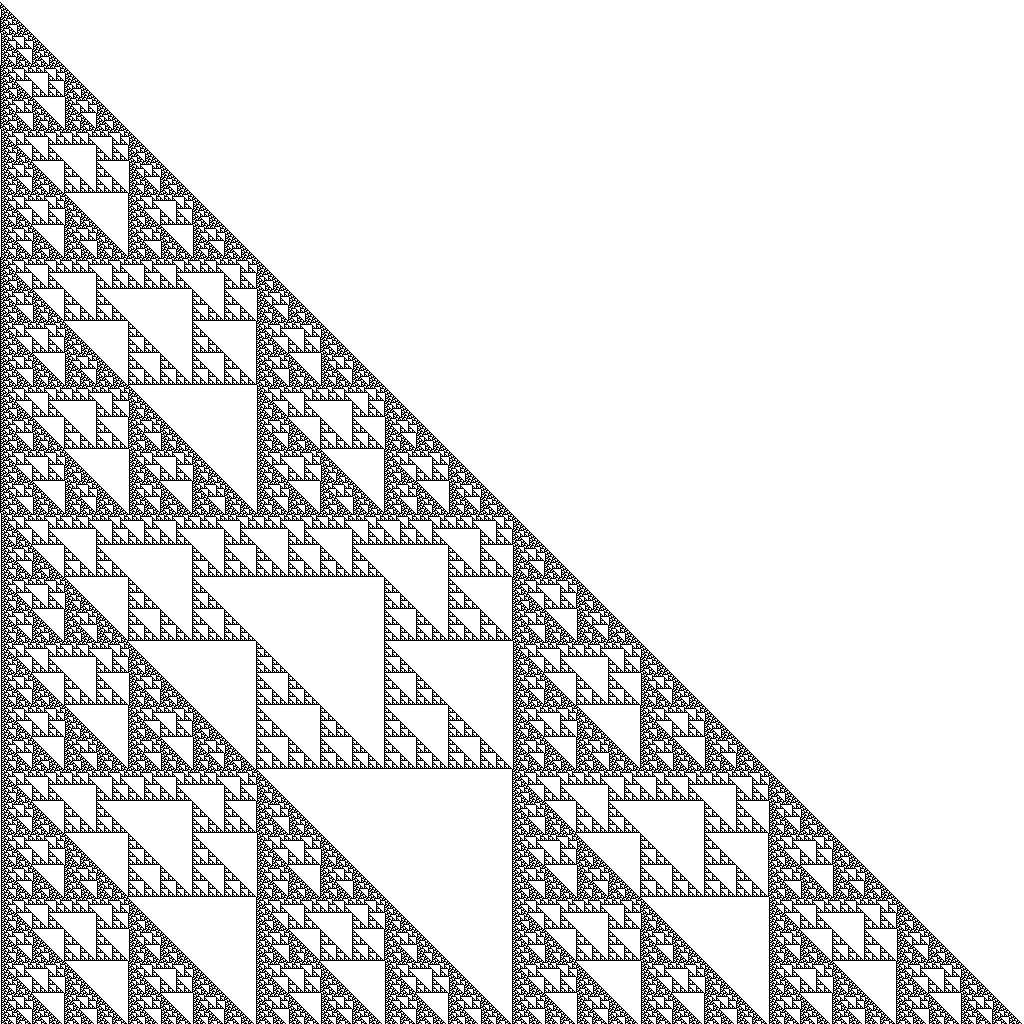}} &
        \subcaptionbox{$[a]a$\label{fig:6}}{\includegraphics[width=0.21\textwidth]{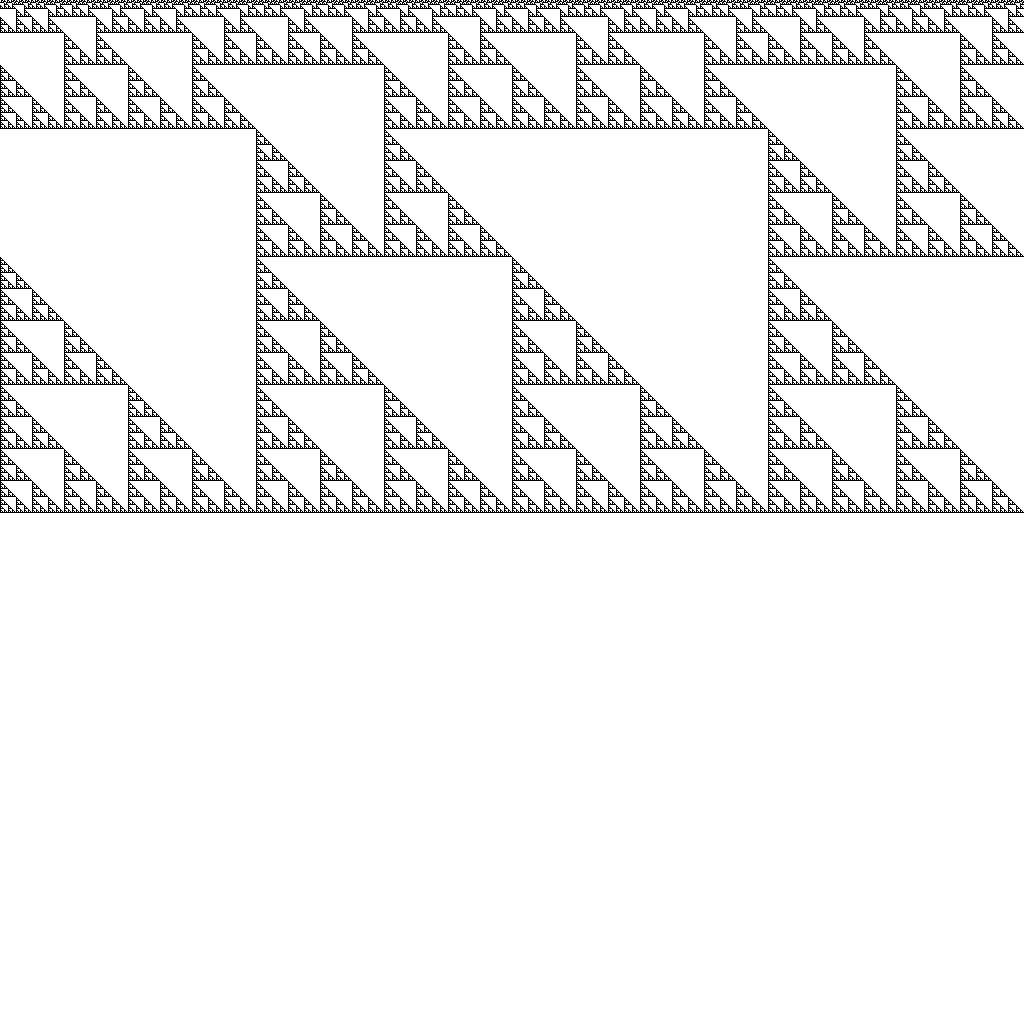}} &
        \subcaptionbox{$[a]b$\label{fig:7}}{\includegraphics[width=0.21\textwidth]{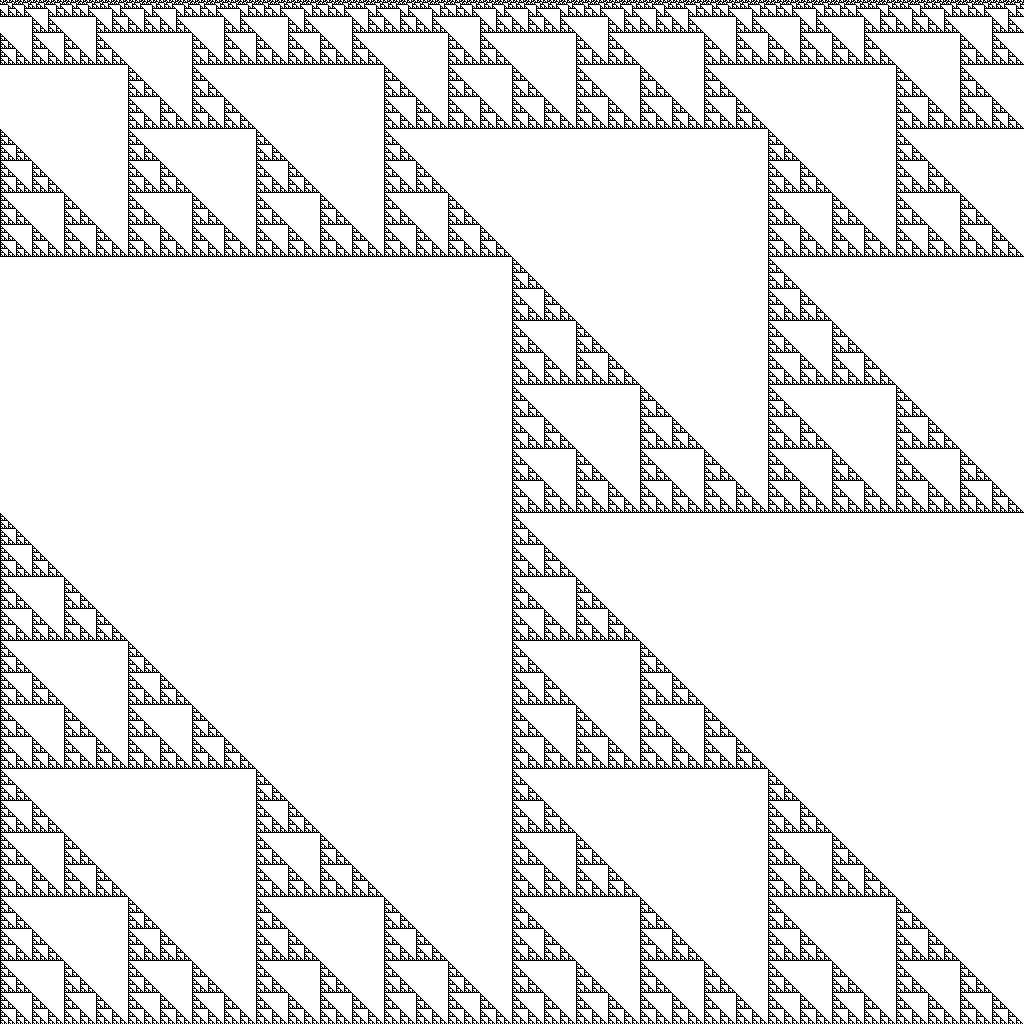}} &
        \subcaptionbox{$[a]ab$\label{fig:8}}{\includegraphics[width=0.21\textwidth]{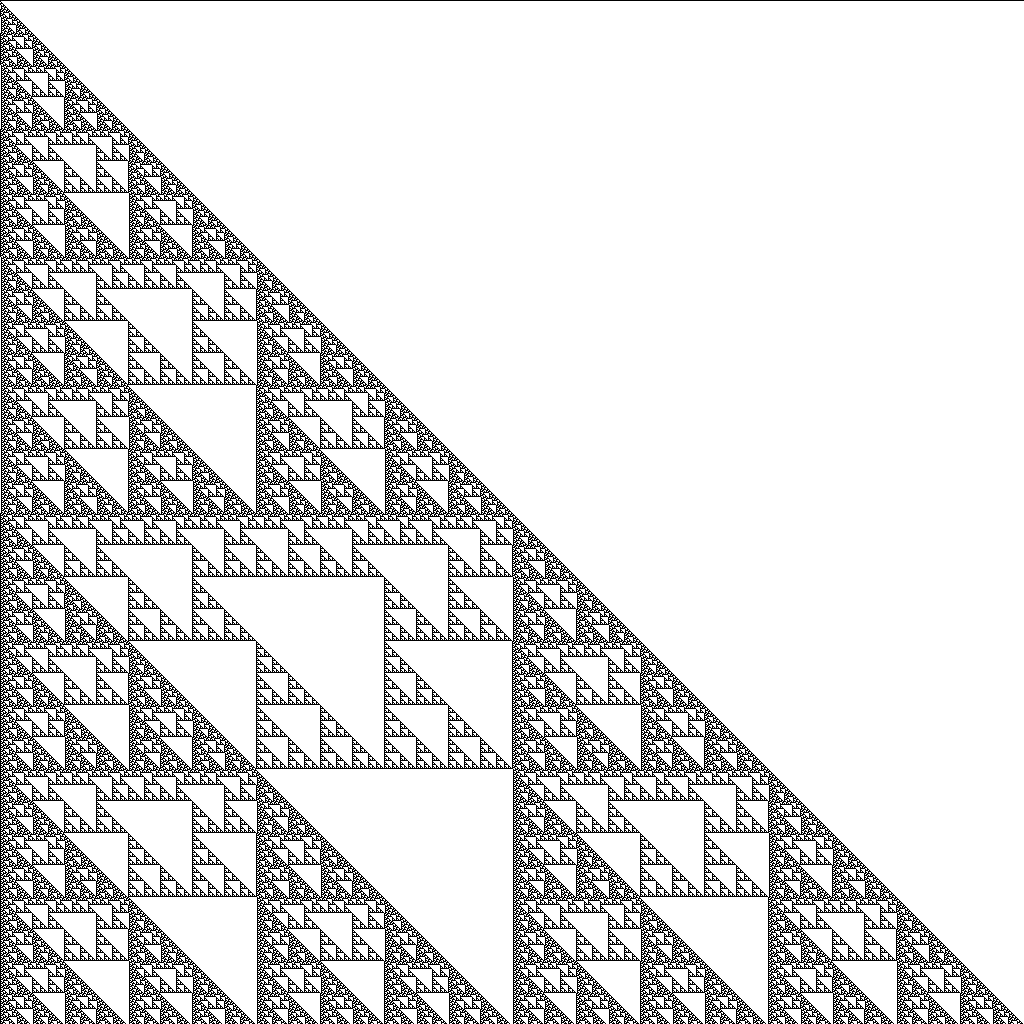}} \\
        
        \subcaptionbox{$[b]$\label{fig:9}}{\includegraphics[width=0.21\textwidth]{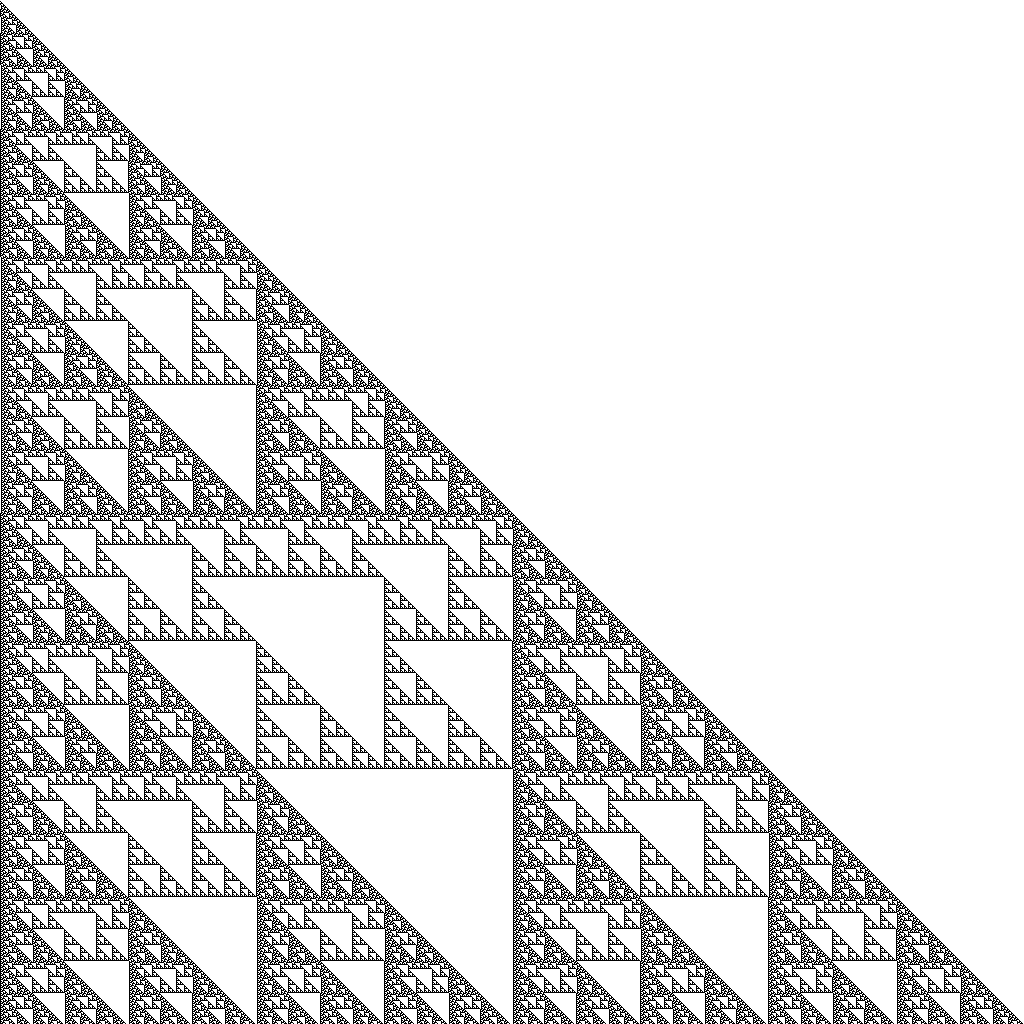}} &
        \subcaptionbox{$[b]a$\label{fig:10}}{\includegraphics[width=0.21\textwidth]{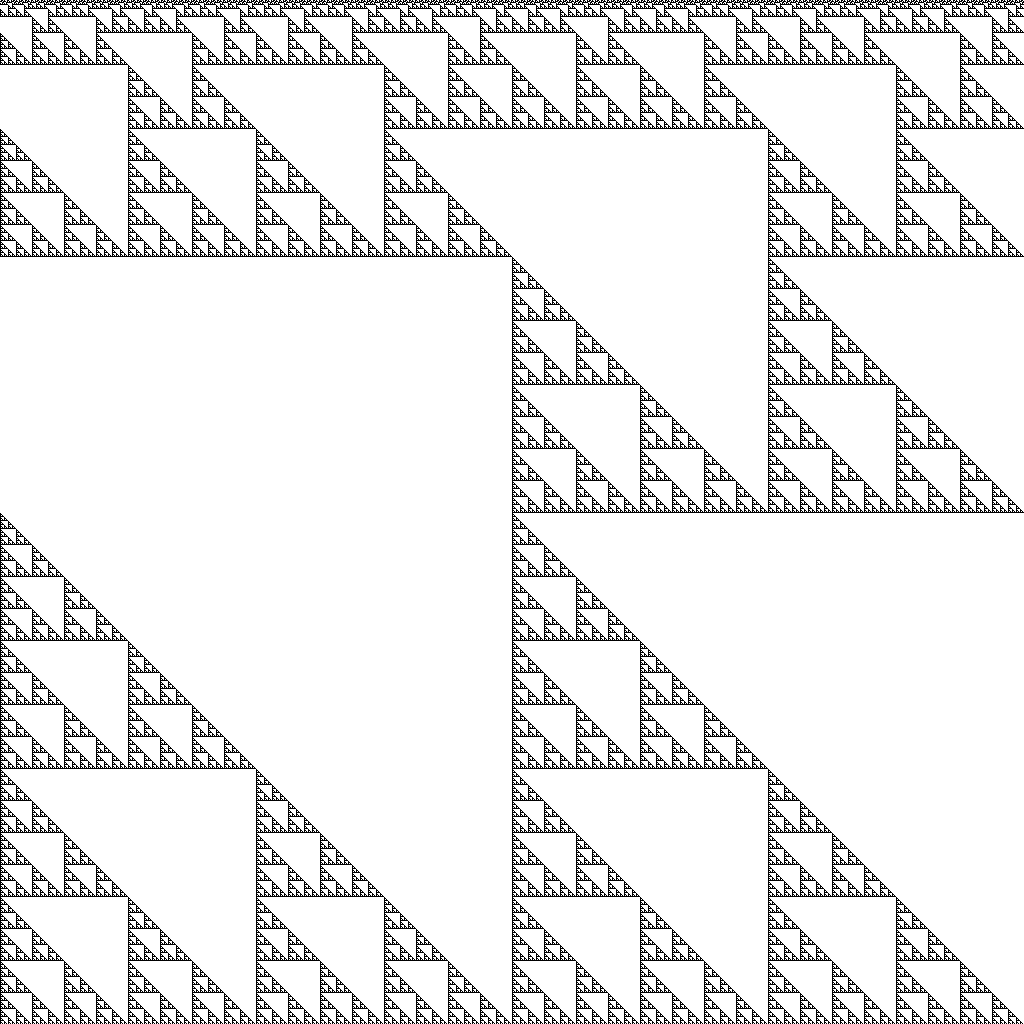}} &
        \subcaptionbox{$[b]b$\label{fig:11}}{\includegraphics[width=0.21\textwidth]{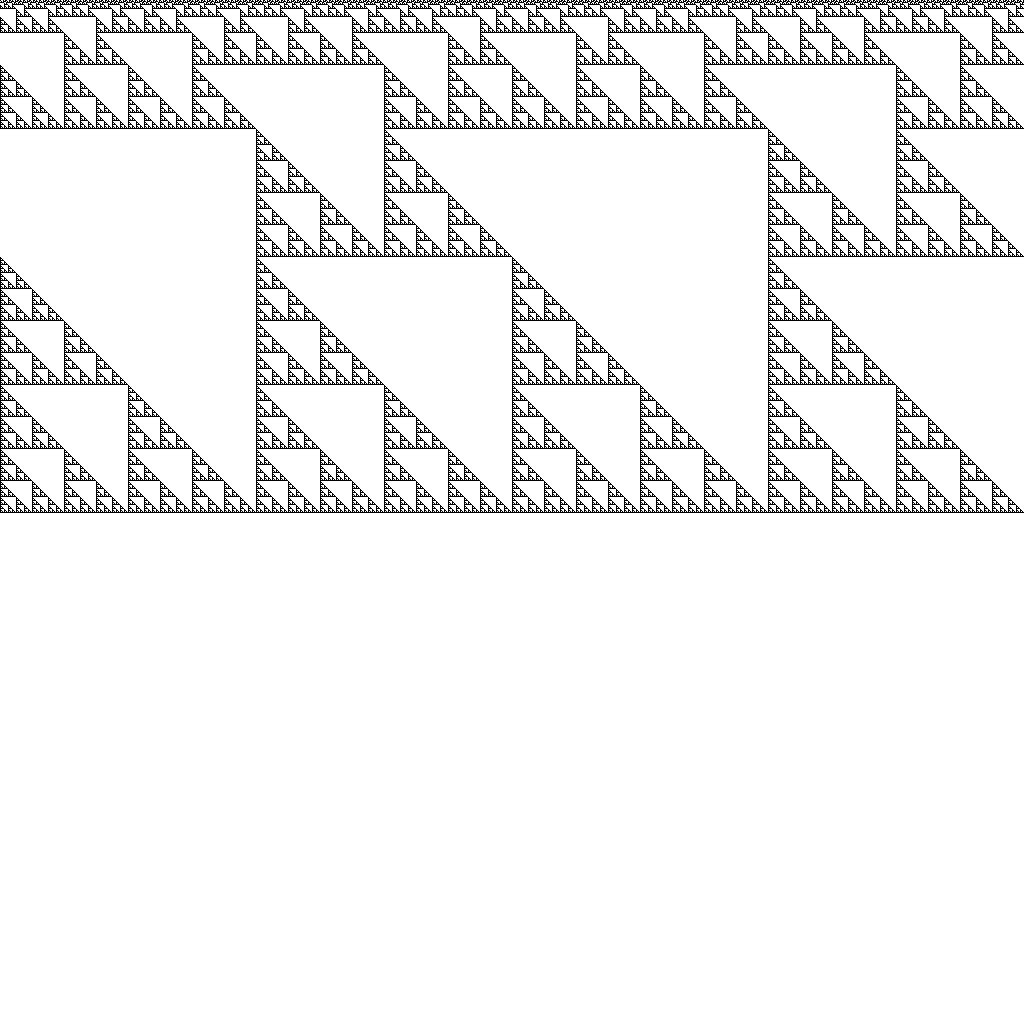}} &
        \subcaptionbox{$[b]ab$\label{fig:12}}{\includegraphics[width=0.21\textwidth]{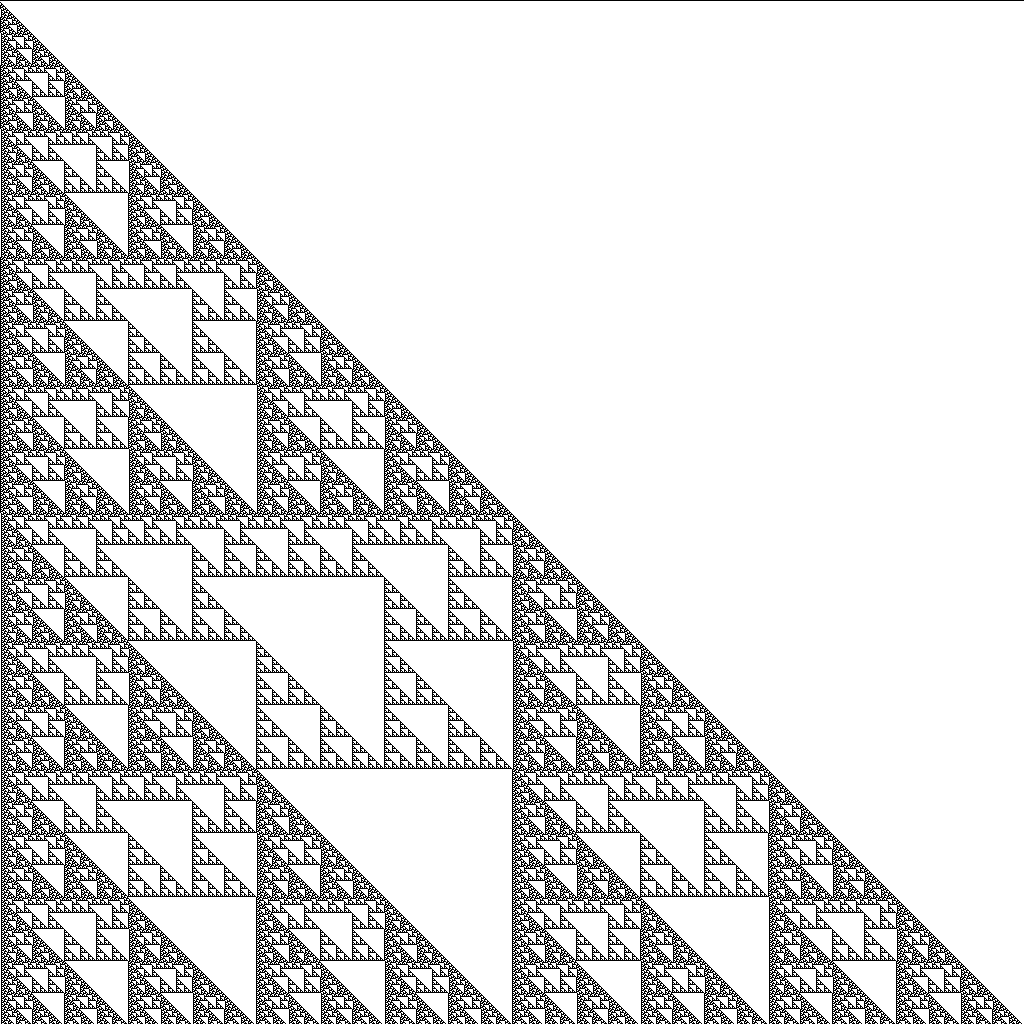}} \\
        
        \subcaptionbox{$[ab]$\label{fig:13}}{\includegraphics[width=0.21\textwidth]{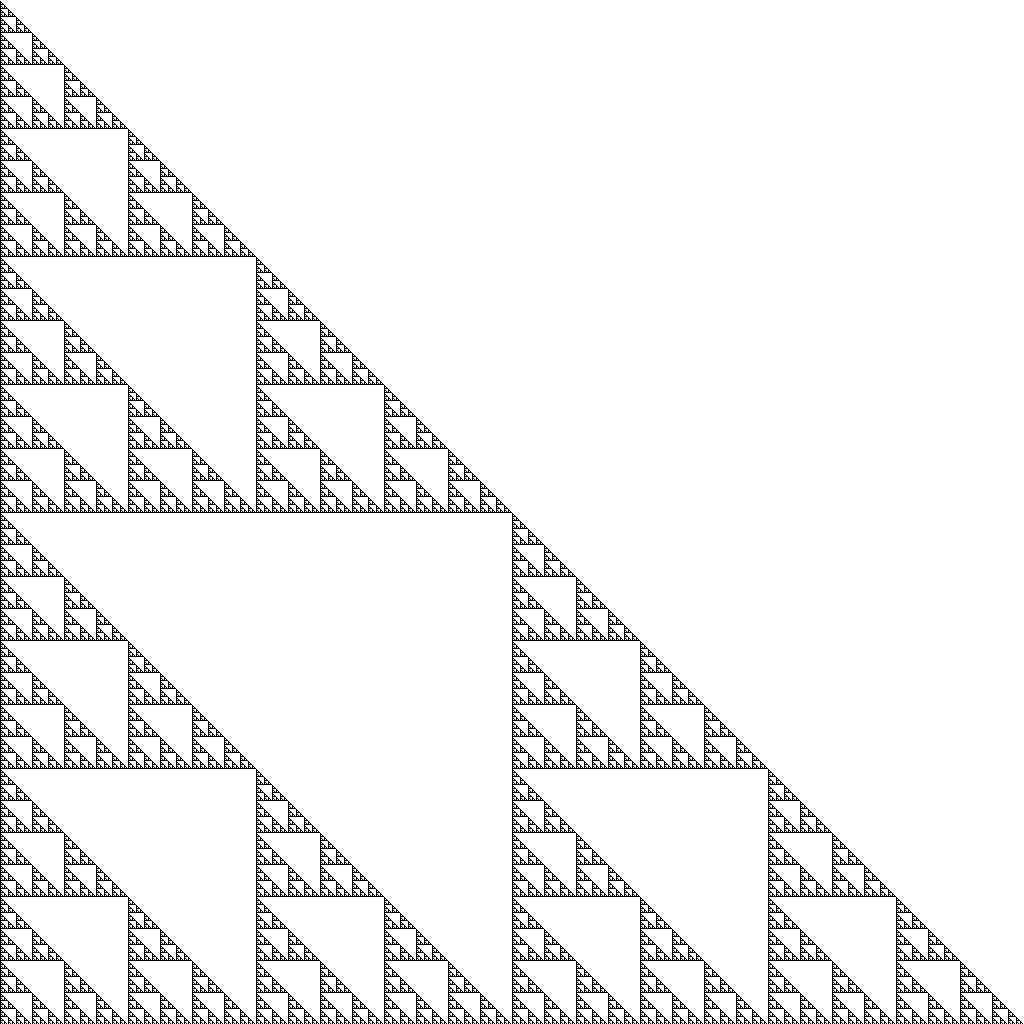}} &
        \subcaptionbox{$[ab]a$\label{fig:14}}{\includegraphics[width=0.21\textwidth]{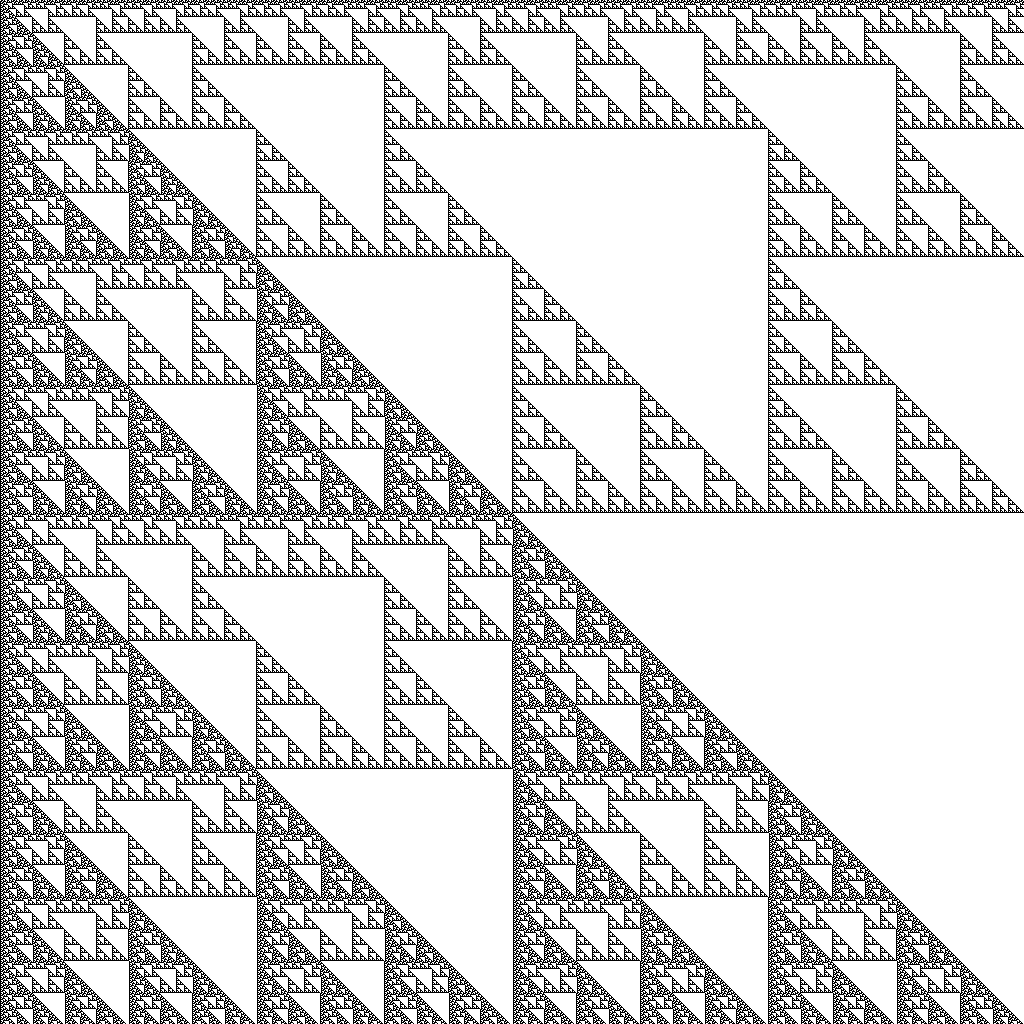}} &
        \subcaptionbox{$[ab]b$\label{fig:15}}{\includegraphics[width=0.21\textwidth]{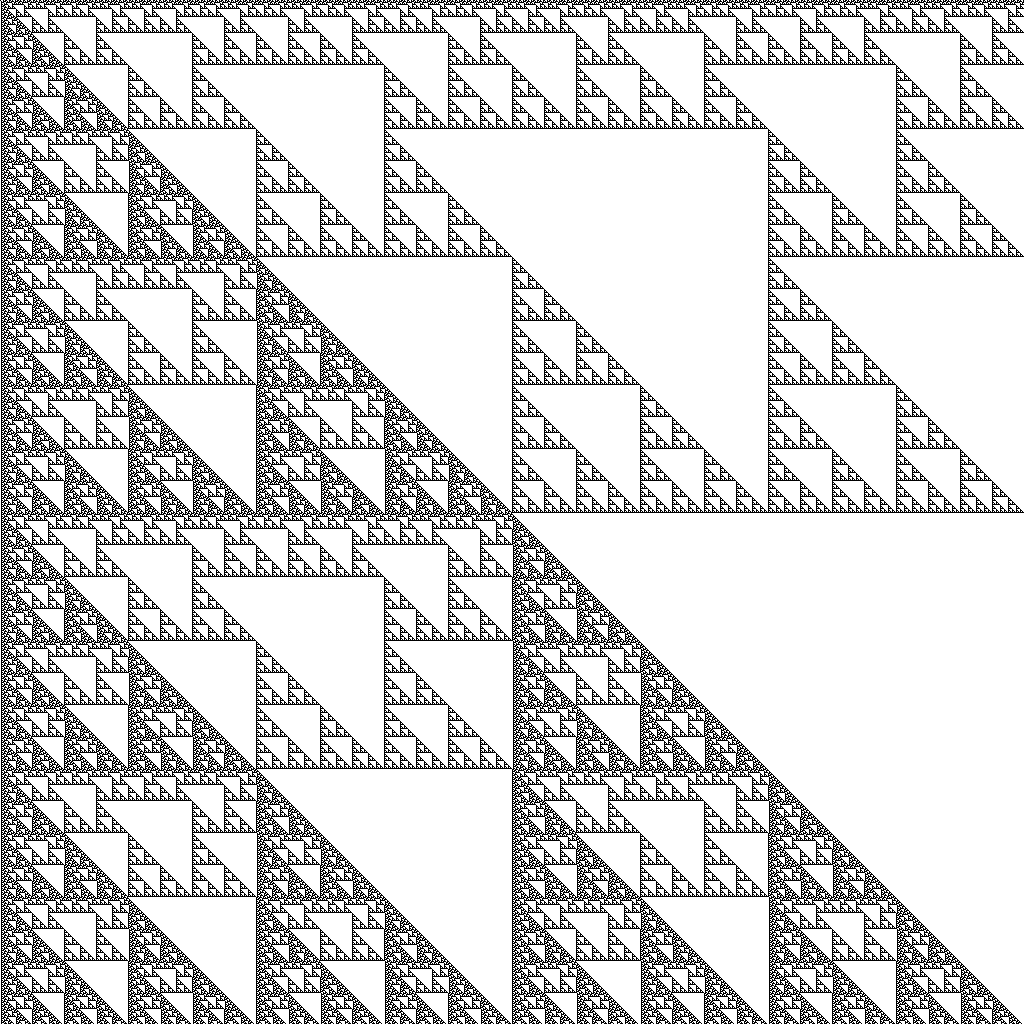}} &
        \subcaptionbox{$[ab]ab$\label{fig:16}}{\includegraphics[width=0.21\textwidth]{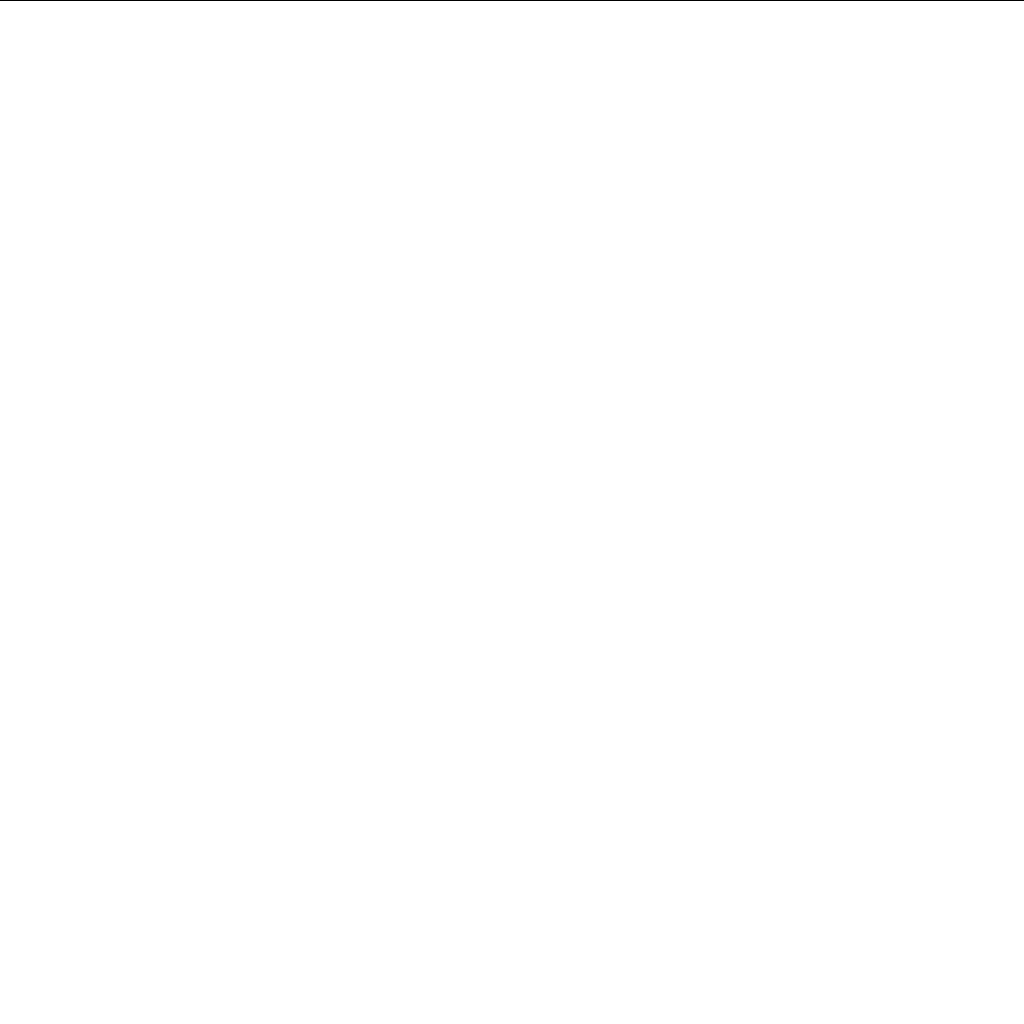}} \\
    \end{tabular}
    \caption{The 16 subpatterns in Figure~\ref{fig:PTM_Ledrappier}.}
    \label{fig:16subpatterns}
\end{figure}

One can also think of the states of the substitution system as subpatterns of Figure~\ref{fig:PTM_Ledrappier}.  The rule  $[]a\rightarrow \begin{array}{|c|c|} \hline []a & [a]b \\ \hline []a & [a]ab \\ \hline \end{array}$ says that the pattern $[]a$ is quite literally a collage of the four patterns $[]a$, $[a]b$, $[]a$ and $[a]ab$.  Figure~\ref{fig:16subpatterns} represents the subpatterns associated to each state of the substitution system.  It proves in a nice way that the substitution system is minimal, since each pattern is different: while some may seem equal at first glance, they are all distinct. For instance, the patterns $[a]$ and $[b]$ appear similar, but their sum modulo 2, $[ab]$, is clearly not null.

Let's recapitulate: starting from a substitution system with state set $E$ for the initial configuration, and a substitution system with state set $F$ for the cellular automaton (more precisely for the Green functions $(x\mapsto F^j(\bar x)_i)_{i,j}$ of the cellular automaton), we constructed a substitution system with state set $(\Z_2)^{(E\times F)^2}$ for the whole spacetime diagram, before cutting down its size. 

The idea underlying the next section is that the same idea works in general when the alphabet $\Sigma$ is a commutative monoid such that $\pi(\Sigma)\subseteq\{p\}$, and the initial configuration is $p$-automatic.  Instead of having formal sums modulo 2, we have formal sums with coefficients in some $\freemonoid{\iindex}{\period}$.
It is a sound idea and works as described; the slightly tricky part is making sure that the sums are finite.

When $c$ is an arbitrary, not necessarily finite, configuration, for every $i\in\Z$ and $j\in\N$, $F^j(c)_i=\prod\limits_k F^j(\overline{c_k})_{i-k}$ is still a finite product, because for every fixed $j$,  almost all Green functions $x \mapsto F^j(\bar x)_i$ are trivial.    The problem is that we do not work directly with the Green functions: we work with the states of a substitution system thereof, for which ``being trivial" doesn't mean anything \emph{per se}.  We have to manually tag that attribute of triviality to the states of the substitution system.  It is not surprising that this can be done, nor is it difficult.  In Section~\ref{sec:automatic_initial_configuration_general_case}, (\ref{eq:property_R}) defines a property $\mathcal{R}$ on the set of states of the substitution system for the Green functions: a state has property $\mathcal{R}$ iff, from this state, a nontrivial Green function can be reached.
A state is thus ``trivial" if it does not fulfill $\mathcal{R}$.   It is then a matter of proving that, for a given $j$, only a finite number of cells of coordinates $(i,j)$ are associated to a state fulfilling $\mathcal{R}$.  Then, what remains to be done is just a generalization of the method employed in our example.

\subsection{General Case}\label{sec:automatic_initial_configuration_general_case}

In order to prove Theorem~\ref{thm:automatic_initial_configuration}, we introduce the following lemma, which states a general condition under which two automata describing $k$-automatic functions can be combined into one.

\begin{lemma}\label{lemma4}
Let $k\geq 2$ and $d\geq 1$ be integers.  Let $X,Y$ be finite sets including respectively the elements $\zero{X}$ and $\zero{Y}$.
Let $e:\Z^{d} \to X$, $f:\Z^{d}\to Y$, and, for $\multi{s}\in\intint{0,k-1}^{d}$, $\epsilon_{\multi{s}}:X\to X$ and $\phi_{\multi{s}}:Y\to Y$  such that
\begin{equation}
\text{$\epsilon_{\multi{s}}(\zero{X}) = \zero{X}$ and $\phi_{\multi{s}}(\zero{Y})=\zero{Y}$} \label{eq:lemma_zero}
\end{equation}
and, for all $\multi{n}\in\Z^{d}$,
\begin{equation}
e(k\multi{n}+\multi{s}) = \epsilon_{\multi{s}} \circ e (\multi{n}) \text{ and } f(k\multi{n}+\multi{s}) = \phi_{\multi{s}} \circ f (\multi{n}). \label{eq:lemma}
\end{equation}

Let $(M,+,0)$ be a finite commutative monoid and $\nu:X\times Y\to M$ such that $\forall (x,y)\in X\times Y\; \nu(x,\zero{Y})=\nu(\zero{X},y)=0$.

Assume that, for every $\multi{n}\in\Z^d$, 
\begin{equation}\label{eq:finite_sum}
\text{$\left \{(\multi{k},\multi{l})\in\Z^d\times\Z^d , \text{$\multi{k}+\multi{l}=\multi{n}$ and $e(\multi{k})\neq\zero{X}$ and $f(\multi{l})\neq\zero{Y}$} \right\}$ is finite.}
\end{equation}
Define $W:\Z^d\to M$ by
$W(\multi{n}) = \sum\limits_{\multi{k}+\multi{l}=\multi{n}}\nu(e(\multi{k}),f(\multi{l}))$.   Then $W$ is $k$-automatic.
\end{lemma}

\begin{proof}
Let $\iindex$ be the maximum of the thresholds, and $\period$ the least common multiple of the periods of the elements of $M$.
Let us denote $X^* = X\setminus\{\zero{X}\}$ and $Y^* =  Y\setminus\{\zero{Y}\}$.
Let $\Sigma$ be the quotient of the free commutative monoid generated by $X\times Y$ by the relations $x^{\iindex+\period}\sim x^{\iindex}$ and $(\zero{X},y)\sim(x,\zero{Y})\sim 0$: 

\begin{equation}
\Sigma \simeq \freemonoid{\iindex}{\period}^{|X^*\times Y^*|}.\label{eq:monstrueux}
\end{equation}

Let us denote $b$ the function that to an element of $X\times Y$ associates its image in $\Sigma$: if $(x,y)\in X^*\times Y^*$,  $b(x,y)$ is the corresponding basis element of $\Sigma$, and $b(x,y)=0$ if $x=\zero{X}$ or $y=\zero{Y}$. For any $\sigma\in\Sigma$, there are unique $\sigma_{x,y} \in \intint{0,\iindex+\period-1}$ such that 

\begin{equation}
\sigma = \sum\limits_{(x,y)\in X^*\times Y^*}   \sigma_{x,y} b(x,y).
\end{equation}

Let $\widetilde{\nu}:\Sigma\to M$ be the morphism defined by $\tilde\nu (b(x,y)) = \nu(x,y)$. It is well defined because, if $b(x,y)=0$, then $x=\zero{X}$ or $y=\zero{Y}$, so $\nu(x,y)=0$.
For $\multi{s},\multi{t}\in \intint{0,k-1}^d$, let $\gamma_{\multi{s},\multi{t}}$ be the endomorphisms of $\Sigma$ defined by  $\gamma_{\multi{s},\multi{t}}(b(x,y))  = b(\epsilon_{\multi{s}}(x),\phi_{\multi{t}}(y))$.   They are well defined because if $b(x,y)=0$, then $b(\epsilon_{\multi{s}}(x),\phi_{\multi{t}}(y))=0$.

Let $g:\Z^d\to \Sigma$ be defined by 
\begin{equation}
g(\multi{n}) = \sum\limits_{ \multi{k} + \multi{l}=\multi{n}}b(e(\multi{k}),f(\multi{l})).
\end{equation}
Notice that, because of (\ref{eq:finite_sum}), the above sum is finite, and therefore well defined.  Moreover, $W=\tilde\nu \circ g$.

Let $\multi{n}\in\Z^d$ and $\multi{s}\in\intint{0,k-1}^{d}$. We have the following:

\begin{align*}
g(k\multi{n}+\multi{s}) & = \sum\limits_{ \multi{k} + \multi{l}=k\multi{n}+\multi{s}}b\left(e(\multi{k}),f(\multi{l})\right)\\
&=\sum\limits_{ k(\multi{q}+\multi{r}) + (\multi{t}+\multi{u})=k\multi{n}+\multi{s}} b \left(e(k\multi{q}+\multi{t}),f(k\multi{r}+\multi{u})\right) \\
&=\sum\limits_{\multi{t},\multi{u}\in \intint{0,k-1}^{d}}\sum\limits_{ \multi{q}+\multi{r} = \multi{n} + \frac{1}{k} (\multi{s} - \multi{t}-\multi{u})} b(\epsilon_{\multi{t}}\circ e(\multi{q}), \phi_{\multi{u}}\circ f(\multi{r}))\\
&=\sum\limits_{\multi{t},\multi{u}\in \intint{0,k-1}^{d}}\sum\limits_{ \multi{q}+\multi{r} = \multi{n} + \frac{1}{k} (\multi{s} - \multi{t}-\multi{u})} \gamma_{\multi{t},\multi{u}} \left(b\left(e(\multi{q}),f(\multi{r})\right) \right) \\
g(k\multi{n}+\multi{s})&=\sum\limits_{\begin{array}{c}\multi{t},\multi{u}\in \intint{0,k-1}^{d}\\\multi{s} - \multi{t}-\multi{u} \in k\Z^d\\ \end{array}} \gamma_{\multi{t},\multi{u}}\circ g \left(\multi{n} + \frac{1}{k} (\multi{s} - \multi{t}-\multi{u}) \right) 
\end{align*}

Therefore, according to Proposition~\ref{prop:grouping}, $g$ is $k$-automatic; and since $W= \tilde\nu\circ g$, so is $W$.
\end{proof}

We can now prove Theorem~\ref{thm:automatic_initial_configuration} and return to multiplicative notation for the monoid operation.  Let $p$ be a prime number and $\Sigma$ a finite commutative monoid  such that $\pi(\Sigma)\subseteq \{p\}$.  Let $F:\Sigma^{\Z}\to \Sigma^{\Z}$ be a cellular automaton that is also an endomorphism of $\Sigma^{\Z}$.  Let $c\in \Sigma^\Z$ be a $p$-automatic configuration.

By definition, there exist a finite set $E$,  functions $d:\Z \to E $ and $\tau: E\to \Sigma$ and, for $s\in\intint{0,p-1}$,  functions $\delta_{s}:E\to E$ such that for all $n\in\Z$, $d(pn+s) = \delta_{s} \circ d (n)$ and $\tau\circ d(n) = c_n$.

Let $X$ be the disjoint union of $E$ and $\zero{X}$. Let us define $e:\Z^2\to X$ by
\begin{equation}
e(i,j)=\left\{  \begin{array}{ll} d(i)& \text{if $j=0$} \\ \zero{X} & \text{if $j\neq 0$} \end{array} \right..
\end{equation}

Notice that, for every $n\in\Z$, $\tau\circ e(n,0)=c_n$.  For $(s,t) \in \intint{0,p-1}^2$, let us define the function $\epsilon_{(s,t)}: X\to X$ by
\begin{equation}
   \epsilon_{(s,t)}(x)=\left\{  \begin{array}{ll} \delta_s(x) & \text{if $x\in E$ and $t=0$} \\ \zero{X} & \text{otherwise} \end{array} \right..
\end{equation}

We can check that the functions $e$ and $\epsilon_{(s,t)}$ thus introduced fulfill the first half of conditions~(\ref{eq:lemma_zero}) and (\ref{eq:lemma}). Let $(i,j)\in\Z^2$ and $(s,t)\in\intint{0,p-1}^2$.
\begin{itemize}
\item If $j=t=0$, then $\epsilon_{(s,t)}\circ e(i,j) = \epsilon_{(s,0)}(d(i)) = \delta_s(d(i))= d(pi+s) = e(pi+s,0) = e(pi+s,pj+t)$.
\item If $j\neq 0$ or $t\neq 0$, then  $\epsilon_{(s,t)}\circ e(i,j)  = \zero{X}=e(pi+s,pj+t)$ because $pj+t\neq 0$.
\end{itemize} 

Therefore, in all cases, we do have $\epsilon_{(s,t)}\circ e(i,j) =  e(pi+s,pj+t)$.  We now move on to defining $Y$, $f$ and the functions $\phi_{(s,t)}$.

According to Theorem~\ref{thm:main}, for every $x\in \Sigma$, the spacetime diagram produced by $F$ on the initial configuration $\bar x$ is $p$-automatic.  Therefore the spacetime diagram of the Green functions of $F$, i.e., the double sequence $(x\mapsto F^j (\bar x)_i )_{(i,j)\in \Z\times \N}$, is itself $p$-automatic.    Since $F$ is a cellular automaton, there is some nonnegative integer $r$ called the radius of the automaton such that, if $|i|>rj$, then the Green function $(x\mapsto F^j(\bar x)_i)$ is the trivial morphism $\underline{1}:x\mapsto 1_\Sigma$.

By definition, there exists  a finite set $E'$,  functions $d':\Z^2 \to E' $ and $\tau': E'\to \Sigma^\Sigma$ and, for $(s,t)\in\intint{0,p-1}^2$,  functions $\delta'_{(s,t)}:E'\to E'$ such that for all $(i,j)\in\Z^2$, $d'(p i + s,pj+t) = \delta'_{(s,t)} \circ d' (i,j)$ and $\tau'\circ d'(i,j) = (x\mapsto F^j(\bar x)_i)$.

For $x\in E'$, we denote
\begin{equation} \mathcal{R}(x) \equiv \exists l\geq 0 \; \exists (s_1,t_1),(s_2,t_2),\ldots,(s_l,t_l)\in \intint{0,p-1}^2\quad \tau'\delta'_{(s_1,t_1)}\delta'_{(s_2,t_2)}\ldots\delta'_{(s_l,t_l)}(x) \neq \underline{1} \label{eq:property_R}
\end{equation}

If one has in mind the definition of $k$-automaticity in terms of finite automata, $\mathcal{R}(x)$ means that, from state $x$,  a state that projects to a nontrivial Green function is reachable.
We have the following property:

\begin{equation}
\forall(i,j)\in\Z\times\N \;\forall (s,t)\in \intint{0,p-1}^2\quad \mathcal{R}(d'(pi+s,pj+t)) \Rightarrow\mathcal{R}(d'(i,j)). \label{eq:prop_R}
\end{equation}

The idea is that we are going to identify all the other states, from which only $\underline{1}$ is reachable, with a unique state $\zero{Y}$.   Let $Y$ be the disjoint union of $E'$ and $\zero{Y}$ and let us define $f:\Z^2\to Y$ by

\begin{equation}
f(i,j)=\left\{  \begin{array}{ll} d'(i,j)& \text{if $j\geq 0$ and $\mathcal{R}(d'(i,j))$} \\ \zero{Y} & \text{otherwise} \end{array} \right..
\end{equation}


For $(s,t) \in \intint{0,p-1}^2$, let us define $\phi_{(s,t)}: Y\to Y$ by 

\begin{equation}
   \phi_{(s,t)}(y)=\left\{  \begin{array}{ll} \delta'_{(s,t)}(y) & \text{if $y\in E'$ and $\mathcal{R}(\delta'_{(s,t)}(y))$} \\ \zero{Y} & \text{otherwise} \end{array} \right..
\end{equation}

The functions $\phi_{(s,t)}$ thus introduced clearly fulfill the second half of condition~(\ref{eq:lemma_zero}).  
Let us check that the functions $f$ and $\phi_{(s,t)}$ also fulfill the second half of condition~(\ref{eq:lemma}). Let $(i,j)\in\Z^2$ and $(s,t)\in\intint{0,p-1}^2$.
\begin{itemize}
\item If $j\geq 0$ and $\mathcal{R}(d'(pi+s,pj+t))$, then according to $(\ref{eq:prop_R})$, we have $\mathcal{R}(d'(i,j))$, so $\phi_{(s,t)}\circ f(i,j) = \phi_{(s,t)}(d'(i,j)) =  \delta'_{(s,t)}(d'(i,j)) = d'(pi+s,pj+t) = f(pi+s,pj+t)$.  
\item If $j\geq 0$, $\mathcal{R}(d'(i,j))$ and $\neg \mathcal{R}(d'(pi+s,pj+t)) $, then  $f(i,j) = d'(i,j)$ and $f(pi+s,pj+t)  = \zero{Y}$ and, since $f(i,j)\in E'$ but $\neg\mathcal{R}(\delta'_{(s,t)}(f(i,j))$,  $\phi_{(s,t)}\circ f(i,j) = \star{Y}$.
\item If $j\geq 0$, $\neg \mathcal{R}(d'(i,j))$ and $\neg \mathcal{R}(d'(pi+s,pj+t)) $, then  $f(i,j) = f(pi+s,pj+t)  = \zero{Y}$ and $\phi_{(s,t)}\circ f (i,j) = \zero{Y}$ because $f(i,j)\not\in E'$.
\item If $j<0$ then $\phi_{(s,t)}\circ f(i,j)  = \zero{Y} = f(pi+s,pj+t)$.
\end{itemize} 

Therefore, in all cases, we do have $\phi_{(s,t)}\circ f(i,j) =  f(pi+s,pj+t)$.

We now have to prove that condition  (\ref{eq:finite_sum}) is fulfilled.
Since $e(i,j)\neq \zero{X}$ implies $j= 0$ and $f(i,j)\neq \zero{Y}$ implies $j\geq 0$, what we have to check is that for every $j\in\N$, 
$\{i\in\Z , f(i,j)\neq\zero{Y} \}$ is finite.

And this is true because, as we have already mentioned, our cellular automaton $F$ has a radius $r$ such that, if $|i|>rj$, then the Green function $(x\mapsto F^j(\bar x)_i)$ is the trivial morphism  $\underline{1}$.
To verify this, observe that for any $x_0\in \Z$, and finite sequence $(s_1,s_2,\ldots,s_l)\in \intint{0,p-1}^l$, if we define, for $i\in\intint{1;l}$, $x_i = px_{i-1}+s_i$,   then $p^l x_0\leq x_l \leq  p^{l}(x_0+1)$.
Now, let $(i,j)\in\Z\times\N$ be such that $\mathcal{R}(d'(i,j))$.   If $i\geq 0$, there must exist a nonnegative integer $l$ such that $p^l  i \leq r p^l(j+1)$, so $i\leq r(j+1)$.
If $i<0$, there must exist a nonnegative integer $l$ such that $p^l |i+1| \leq r p^l(j+1)$, so $|i+1|\leq r(j+1)$.  So, for any given $j\in\N$, $\{i\in\Z , f(i,j)\neq\zero{Y} \} $ is indeed finite.

Let now $\nu:X\times Y\to  \Sigma$ be the function defined by 
\begin{equation}
\nu(x,y) = \left\{  \begin{array}{ll} 1_\Sigma & \text{if $x=\zero{X}$ or $y=\zero{Y}$} \\ \tau'(y)(\tau(x))  & \text{otherwise} \end{array} \right..
\end{equation}

This makes sense because, when $x\neq \zero{X}$ and $y\neq \zero{Y}$, then $\tau(x)\in \Sigma$ and $\tau'(y)\in\Sigma^\Sigma$.  Moreover, for any $i,j,k,l\in\Z$,
\begin{equation}
\nu(e(i,j),f(k,l)) = \left\{ \begin{array}{ll}
1_\Sigma & \text{if $j\neq 0$ or $l<0$ or $\neg\mathcal{R}(d'(k,l))$}\\
F^l(\overline{c_i})_k & \text {otherwise} \\
  \end{array}  \right.
\end{equation}

But since $\neg\mathcal{R}(d'(k,l))$ implies $F^l(\overline{c_i})_k=1_\Sigma$, we have,
\begin{equation}
\text{for every $i,k\in\Z$ and $l\in\N$,  $\nu(e(i,0),f(k,l)) = F^l(\overline{c_i})_k$.}
\end{equation}

And since $F$ is translation invariant,  for any $(k,l)\in\Z\times\N$, we have
\begin{align*}
F^l(c)_k & = \prod\limits_{i\in\Z}  F^l(\overline{c_i})_{k-i} \\
&= \prod\limits_{i+j=k} \nu(e(i,0),f(j,l))\\
& = \prod\limits_{(i,m)+(j,n) = (k,l)}\nu(e(i,m),f(j,n)) \\
\end{align*}

We recognize $W(k,l)$, where $W$ is the function defined in Lemma~\ref{lemma4}. The spacetime diagram of $F$, starting on the initial configuration $c$, is therefore, according to this lemma, $p$-automatic.

\section*{Conclusion}

It is perhaps worth mentioning a few things.

\begin{itemize}
\item It is possible to separate a spacetime diagram into its $p$-automatic components, simply by writing the group $\Z_\period^r$ from Section~\ref{sec:free_monoids} as the product of its Sylow $p$-subgroups.  

\item The proof of Theorem~\ref{thm:main} is constructive and yields, in principle, an algorithm that, from descriptions of $(\Sigma,\cdot)$,  the transition rule and the initial configuration, produces a description of the spacetime diagram in terms of $k$-automatic sequences.  Working out the details of this algorithm would be tedious, and for now not very useful, as its complexity would likely be prohibitive.  The substitution systems derived in \cite{Gutschow2010fractal} were already quite large, and the proof of Proposition~\ref{prop:aperiodic} contains a finite automaton that has $1+|\Sigma|^{\omega}\times 2^{\frac{\omega(\omega-1)}{2}}$ states, with an alphabet of size $|\supp(c)|^{\omega} +|I|^\omega$, which, considering $\omega$ may have to be chosen at least as large as $|M|-1$ --- as in the case of $P_n$ in Section~\ref{sec:deuxieme_exemple} --- is impractical.

\item In this article, the grid of the cellular automaton is the one-dimensional $\Z$, but the results should extend to grids $\Z^d$ for $d\geq 2$. The only obstacle must be the proliferation in notation, which is already cumbersome in dimension~1.

\item In the statement of Theorem~\ref{thm:main}, $\pi(\Sigma)$ is always a set of primes: That in itself is a bit puzzling.

\item It is tempting to imagine that Theorem~\ref{thm:automatic_initial_configuration} can be generalized in the following way: "for any finite commutative monoid $\Sigma$, if the initial configuration is $A$-automatic, then the spacetime diagram is $\pi(\Sigma)\cup A$-automatic".  That doesn't seem to be true, though. With the rule defining Pascal's triangle modulo 2, that is, with $\Sigma=\Z_2$, if the initial configuration $c$ is defined by $c_n = \left\{ \begin{array}{ll} 1 & \text{if $n$ is a power of 3}\\0&\text{otherwise} \end{array}\right.$, it is not clear that the spacetime diagram is $\{2,3\}$-automatic, or even that the problem of calculating the state of cell $(i,j)$ has particularly low complexity (see Figure~\ref{fig:powersof3Sierpinski}).

\begin{figure}[htbp]
\centering
\includegraphics[width=0.8\textwidth]{powersof3Sierpinski.png}
\caption{The spacetime diagram on $\intint{0,2^{13}-1}^2$ of Pascal's triangle modulo 2 with the powers of 3 as initial configuration.\label{fig:powersof3Sierpinski}}
\end{figure}

\item Can it be proven, in the spirit of Cobham-Sem\"enov theorem (\cite{semenov1977presburger}), that if a double sequence is both $A$-automatic and $B$-automatic, where $A$ and $B$ are both sets of primes, then it must be $(A\cap B)$-automatic?
More generally, it would be useful to devise a way of disproving $A$-automaticity, perhaps by defining something like an $A$-kernel.

\item 
Lastly, let us add that it feels like Proposition~\ref{prop:corrigee}, and/or Theorem~\ref{thm:GNW10}, should be an easy consequence of some generalization of Christol's and Salon's theorems (\cite{Christol1979ensembles} and \cite{salon1986suites}), although it is not yet clear to the author how this would work.

\end{itemize}
\acknowledgements
We  thank the anonymous reviewer for their patient and careful reading of the many versions of this paper. 

\bibliographystyle{abbrvnat}
\bibliography{biblio}
\end{document}